\documentclass[aps,prl,twocolumn,superscriptaddress,floatfix,showpacs]{revtex4}
\usepackage{textcomp}
\usepackage{amsmath}
\usepackage{amssymb}
\usepackage{graphicx,graphics}
\usepackage{latexsym,verbatim}
\usepackage{dcolumn} 
\usepackage{color}
\usepackage{ulem}
\usepackage[latin1]{inputenc}

\begin{document}

\title{Time-reversal and rotation symmetry breaking superconductivity in Dirac materials}
\author{Luca Chirolli}
\thanks{luca.chirolli@imdea.org}
\affiliation{IMDEA-Nanoscience, Calle de Faraday 9, 
E-28049 Madrid, Spain}
\author{Fernando de Juan}
\affiliation{IMDEA-Nanoscience, Calle de Faraday 9, 
E-28049 Madrid, Spain}
\affiliation{Rudolf Peierls Centre for Theoretical Physics, 
Oxford, 1 Keble Road, OX1 3NP, United Kingdom}
\author{Francisco Guinea}
\affiliation{IMDEA-Nanoscience, Calle de Faraday 9, 
E-28049 Madrid, Spain}
\affiliation{Department of Physics and Astronomy, 
University of Manchester,
Oxford Road, Manchester M13 9PL, United Kingdom}

\begin{abstract}
We consider mixed symmetry superconducting phases in Dirac 
materials in the odd parity channel, where pseudoscalar and 
vector order parameters can coexist due to their similar 
critical temperatures when attractive interactions are of 
finite range. We show that the coupling of these order 
parameters to unordered magnetic dopants favors the 
condensation of novel time-reversal symmetry breaking (TRSB) 
phases, characterized by a condensate magnetization, 
rotation symmetry breaking, and simultaneous ordering of the 
dopant moments. We find a rich phase diagram of mixed TRSB 
phases characterized by peculiar bulk quasiparticles, with 
Weyl nodes and nodal lines, and distinctive surface states. 
These findings are consistent with recent experiments on 
Nb$_x$Bi$_2$Se$_3$ that report evidence of point nodes, 
nematicity, and TRSB superconductivity induced by Nb 
magnetic moments.
\end{abstract}


\maketitle

{\it Introduction --}   
One of the most fascinating aspects of unconventional 
superconductivity is that the condensate can display 
spontaneous time reversal symmetry breaking (TRSB), 
hosting an intrinsic Cooper pair 
magnetization~\cite{SigristUeda,Maeno}. This can occur only 
with a multicomponent order parameter when the different 
components develop relative phases, as in the well known 
$p+ip$ chiral state proposed for Sr$_2$RuO$_4$ or the $d+id$ 
state conjectured for some cuprate superconductors 
\cite{SigristUeda}. Experimental evidence of 
TRSB superconductivity has been obtained from muon spin 
rotation $\mu$SR in UPt$_3$ \cite{Luke1993} and 
Sr$_2$RuO$_4$ \cite{Luke1998}, from the polar Kerr 
effect~\cite{Kapitulnik} and from Josephson tunneling 
experiments. The two-dimensional $p+ip$ state in particular 
has attracted great interest as a topological superconductor 
with protected edge and vortex modes, of potential use in 
the field of quantum computation \cite{Nayak,Qi}. In a three 
dimensions chiral SC is also possible, allowing the 
realization of a Weyl superconductor with Majorana arcs on 
the surface \cite{Meng2012,Sau2012,Yang2014}, but realistic 
candidate materials for this superconducting state are 
lacking.    

Recently, very compelling evidence for unconventional 
superconductivity has been reported in Dirac materials of 
the Bi$_2$Se$_3$ family upon doping 
\cite{Zhang-NP2009,Hasan2010,Qi}. These studies were originally 
motivated by the prediction of a three dimensional, time-
reversal invariant (TRI) topological superconductor 
featuring protected Andreev surface states~\cite{FuBerg}. 
However, the rich phenomenology gathered so far suggests a 
more complicated pairing scenario. Superconductivity was 
first observed in Cu$_x$Bi$_2$Se$_3$ 
\cite{Hor-PRL2010,Wray-NP2010,Kriener-PRL2011}, but evidence for the characteristic 
surface Andreev states has remained 
controversial
~\cite{Sasaki-PRL2011,Levy-PRL2013,Peng-PRB2013}. Moreover, nuclear magnetic resonance experiments 
~\cite{Matano} reveal that there is spin rotation 
symmetry breaking in the superconducting state, which rather 
supports a different pairing state of nematic type +
~\cite{FuPRBR2014,Venderbos2016}. Superconductivity was also 
reported in Sr$_x$Bi$_2$Se$_3$~\cite{Shruti,Liu} and in 
Tl$_x$Bi$_2$Te$_3$~\cite{Wang}, but evidence for 
unconventional pairing is lacking. Most interestingly, 
superconductivity has also been reported in 
Nb$_x$Bi$_2$Se$_3$~\cite{Qiu}, where initially paramagnetic 
samples were shown to develop a spontaneous magnetization at 
the superconducting transition. The magnetization survived 
only at the surface in the Meissner state, and it was 
claimed to originate from Nb magnetic moments. In the same 
compound, a later torque magnetometry experiment 
~\cite{Asaba} showed clear signatures of rotation symmetry 
breaking, and penetration depth measurements revealed a 
power law dependence with temperature~\cite{Smylie} which 
points to the existence of nodes in the gap.

This complicated phenomenology is perhaps best understood 
within the minimal model of a superconducting Dirac 
Hamiltonian with approximate rotation symmetry, where there 
are only three possible pairing channels: a conventional 
$s$-wave scalar, an odd-parity pseudoscalar, and a vector. 
The pseudoscalar order parameter $\chi$ corresponds to the 
TRI topological superconductor, while rotation symmetry 
breaking can only be produced by the vector
$\boldsymbol{\psi}$. The condensation of $\boldsymbol{\psi}$ 
is therefore a prerequisite to explain current experiments, 
but it has previously been shown that with only local 
interactions the $\chi$ channel always has a higher critical 
temperature than the $\boldsymbol{\psi}$ channel 
\cite{FuBerg}. In addition, even if $\chi$ could be ignored, 
$\boldsymbol{\psi}$ remains time-reversal symmetric within 
current models~\cite{FuPRBR2014,Venderbos2015}. These two 
problems make the explanation of the observed phenomenology 
a theoretical challenge.

Motivated by the recent experiments, in this work we develop 
a theory of possible TRSB superconducting phases of doped 
Dirac Hamiltonians in the presence of magnetic impurities. 
We first show that when further neighbor electron-electron 
interactions are included, the critical temperature of 
$\boldsymbol{\psi}$ raises and can become comparable to that 
of $\chi$, providing a solution to the first problem. The closeness of the critical temperatures enables new mixed symmetry phases where both order parameters can condense simultaneously, similar to $s+id$ states predicted in high-T$_c$ superconductors~\cite{sid1,sid2,sid3}. We then develop a theory for these mixed phases, showing that the coupling of magnetic impurities, which would otherwise be paramagnetic, to the magnetization of the Cooper pairs \cite{WalkerPRL2002,Mineev2002,WalkerPRB2002,Mineev2004} favors the condensation of TRSB phases and the consequent ordering of the magnetic impurities. We find three novel mixed TRSB phases that differ in the way rotation and gauge symmetries are broken and can be distinguished by their bulk spectrum, which may be gapped or feature Weyl nodes or nodal lines, or by the existence of surface states. We find a phase that is consistent with the surface magnetization~\cite{Qiu}, rotation symmetry breaking~\cite{Asaba} and the existence of linear nodes \cite{Smylie}.  

{\it Superconductivity in Dirac materials --} We now consider the possible superconducting instabilities of Dirac Hamiltonians. 
To make contact with previous work, we start with the Hamiltonian commonly employed to describe Bi$_2$Se$_3$~\cite{FuBerg}
\begin{equation}\label{H0}
{\cal H}_0=m\sigma_x+v\sigma_z(k_xs_y-k_ys_x)+v_zk_z\sigma_y,
\end{equation} 
where $s_i$ are spin Pauli matrices and $\sigma_i$ are Pauli matrices for $p_z$-orbitals in the top and bottom layer of the quintuple 
layer QL Bi$_2$Se$_3$ structure, $v$ is the Fermi velocity, $m$ the insulating mass.  The time reversal operator is 
${\cal T} = is_y K$ with $K$ complex conjugation. When $v_z=v$, this Hamiltonian is a particular realization 
of the isotropic Dirac Hamiltonian of the form 
\begin{equation}
{\cal H}_0 = \gamma_0 m + v\gamma_0 \gamma_i k_i
\end{equation}
where the Euclidean gamma matrices $\gamma_\mu = (\gamma_0,\gamma_i)$ satisfy $[\gamma_\mu,\gamma_\nu]_+ = 
\mathcal{I}_{\mu\nu}$ and are given by $\gamma_\mu = (\sigma_x,-\sigma_y s_y,\sigma_y s_x,\sigma_z)$. In this work we will 
preferentially use the general Dirac matrices to emphasize the structure of the rotation group: $\gamma_i$ transforms as a vector, 
$\gamma_0$ as a scalar, and the matrix $\gamma_5 \equiv \gamma_0\gamma_1\gamma_2\gamma_3$ as a pseudoscalar.

To classify the possible pairing channels, we introduce the Nambu spinor $\Psi_{\bf k}=({\bf c}^\dag_{\bf k}, i s_y{\bf c}_{-{\bf k}})^T$, 
with ${\bf c}_{\bf k}$ fermionic annihilation operators of ${\cal H}_0$, and consider the Bogolyubov-deGennes Hamiltonian 
$\hat{H}=\frac{1}{2}\int d{\bf k}\Psi^\dag_{\bf k} {\cal H}_{\bf k} \Psi_{\bf k}$, with
\begin{equation}\label{Eq:BdG}
{\cal H}_{\bf k} = ({\cal H}_0({\bf k})-\mu)\tau_z + \Delta_{\bf k} \tau_+ + \Delta^\dag_{\bf k} \tau_-,
\end{equation}
where $\mu$ is the chemical potential,  $\Delta_{\bf k}$ stands for generic momentum-dependent $4\times 4$ pairing matrices and $\tau_i$ 
Pauli matrices act in the particle-hole space. The Nambu construction imposes the charge conjugation symmetry ${\cal C}$ implemented as 
$U_{\cal C} {\cal H}(-{\bf k})^* U_{\cal C}^\dagger = -{\cal H}({\bf k})$, with $U_{\cal C} = s_y \tau_y$, which amounts to the restriction 
$s_y\Delta^*(-{\bf k})s_y = \Delta({\bf k})$. If pairing is momentum independent \cite{FuBerg,Hashimoto}, only six possible matrices in 
the Dirac algebra satisfy this constraint: the two even-parity scalars $I$ and $\gamma^0$, the pseudo-scalar $\gamma^5$ and the vector 
$\gamma^i$, which are both odd under parity. Disregarding the even-parity scalars, the pairing matrix takes the form 
$\Delta=\chi \gamma^5+\boldsymbol{\psi} \cdot \boldsymbol{\gamma}$. For the specific model of Bi$_2$Se$_3$, it was concluded that 
the local interorbital interaction $V$ can give rise to pairing in both of these channels, but the critical temperatures of the two channels satisfy 
$T_\chi \gg T_\psi$, \cite{FuBerg}, which makes it unlikely for the system to condense in the vector 
channel as stated previously. 

We suggest that this problem can be solved by considering momentum-dependent corrections to the two-body interorbital 
density-density interaction. At lowest order in ${\bf q}={\bf k}-{\bf k}'$ one has
\begin{equation}\label{Vpair}
V({\bf k},{\bf k}')=V\left(1+a^2{\bf k}\cdot{\bf k}'\right),
\end{equation} 
with $a$ a length scale on order of the lattice constant. In order to decouple the additional momentum-dependent interaction term 
we need to consider the other ten matrices in the Dirac algebra \cite{SuppMat}. In particular, we note that that pairing matrix 
$\gamma^5\gamma^i k^j\epsilon_{ijk}$ is also a vector, and it modifies the gap matrix as
\begin{equation}
\Delta_{\bf k}=\chi \gamma^5+\boldsymbol{\psi} \cdot (\boldsymbol{\gamma}-ia\gamma^5\boldsymbol{\gamma}\times{\bf k}).
\end{equation}
It is instructive to project the $4\times4$ Dirac matrices into the $2\times2$ space of the Kramers degenerate conduction band 
states relevant to pairing \cite{Venderbos2015}. If we define Pauli matrices $\tilde{s}_i$ for this space, the gap matrix 
takes the form $\Delta_{\bf k}=\chi\tilde{\bf k}\cdot\tilde{\bf s}+\boldsymbol{\psi}\times\tilde{\bf k}\cdot\tilde{\bf s}(1+\mu a/v)$, with 
$\tilde{\bf k}=v{\bf k}/\mu$. Thus, while seemingly of higher order in the Dirac Hamiltonian, the correction term is actually of the same 
order when projected to the Fermi surface. The momentum dependence of the pairing interaction affects only the vector channel and 
it raises its critical temperature $T_{\psi}$, which becomes comparable to $T_{\chi}$ \cite{SuppMat}.

{\it Ginzburg-Landau free energy --}
We now consider superconductivity at the level of the Ginzburg-Landau (GL) free energy. The pseudoscalar order parameter free energy is 
\begin{equation}
F_\chi = a_1 |\chi|^2 + b_1 |\chi|^4 
\end{equation}
and condensation of $\chi$ takes place when $a_1(T_\chi)=0$. For the vector order parameter 
$\boldsymbol{\psi}$, symmetry dictates that the form of the free energy be \cite{Ueda,Knigavko}
\begin{equation}\label{Eq:GLpsi}
F_\psi = a_2 |\boldsymbol{\psi}|^2 + b_2 |\boldsymbol{\psi} |^4  + b_2' |\boldsymbol{\psi} \times \boldsymbol{\psi} ^*|^2.
\end{equation}
The vector representation admits two possible superconducting states: a nematic state $\boldsymbol{\psi} \propto (1,0,0)$ 
which is time-reversal invariant, and a chiral TRSB state $\boldsymbol{\psi} \propto (1,\pm i,0)$ 
\cite{FuPRBR2014,Venderbos2015}. The sign of the coupling $b_2'$ determines whether the vector representation chooses 
the nematic (for $b_2'>0$) or the chiral state (for $b_2'<0$). Since at second order no coupling is allowed by symmetry  
between $\chi$ and $\boldsymbol{\psi}$, the condensation of $\boldsymbol{\psi}$ takes place when 
$a_2(T_\psi)=0$. However, our previous argument suggesting that $a_1\sim a_2$ \cite{SuppMat} and $T_{\chi} \sim T_{\psi}$ 
require that we study a coupled theory beyond second order where both order parameters may coexist. At fourth order the 
coupling term in the GL free energy reads,
\begin{equation}\label{PsiChi-coupling}
F_{\psi,\chi} = d_1 |\chi|^2 |\boldsymbol{\psi}|^2  + d_2 \left|\chi^* \boldsymbol{\psi} 
- \chi \boldsymbol{\psi}^*\right|^2,
\end{equation}
and the total free energy is
\begin{eqnarray}\label{Fchipsi}
F &=& F_\chi+F_\psi+F_{\chi,\psi}.
\end{eqnarray}
In the weak coupling regime with $a_1\sim a_2$ both order parameters acquire a finite value.

The possible TRSB phases arising from this free energy are characterized by a magnetization of the condensate, due to the spin triplet state of the Cooper pairs. By symmetry, the magnetization must be built with gauge invariant combinations of order parameters and transform as a spin, i.e. as a ${\cal T}$-odd pseudo-vector 
(even under inversion). Since $\boldsymbol{\psi}$ is a vector and $\chi$ a pseudoscalar, the following 
combinations satisfy the symmetry requirements,
\begin{equation}
\boldsymbol{\Sigma}_1 = \chi \boldsymbol{\psi}^* - \chi^*\boldsymbol{\psi}, 
\qquad \boldsymbol{\Sigma}_2 = \boldsymbol{\psi}\times \boldsymbol{\psi}^*\nonumber.
\end{equation}
Note that $\boldsymbol{\Sigma}_1$ cannot be built with a standard $s$-wave order parameter because the combination would 
not be a pseudovector. These two pseudovectors are orthogonal and appear quadratically in the GL Eqs.~(\ref{Eq:GLpsi}, 
\ref{PsiChi-coupling}). 


The different possible phases obtained from the GL free energy Eq.~(\ref{Fchipsi}) are realized with different signs of the interaction parameters $b_2'$ and $d_2$ and can be distinguished by the values of ${\bf \Sigma}_1$ and ${\bf \Sigma}_2$ and the way rotation and gauge symmetries are broken. For $d_2, b_2'>0$ one has ${\bf \Sigma}_1={\bf \Sigma}_2=0$ and the system is in the TRI 
nematic phase, with rotation symmetry about the nematic director. When $d_2<0$ and $b_2'>0$ one has ${\bf \Sigma}_1\neq 0$ 
and ${\bf \Sigma}_2=0$, and the system is invariant under rotations about ${\bf \Sigma}_1$. We name this phase TRSB 1. When 
$d_2,b_2'<0$ one has ${\bf \Sigma}_2\neq 0$, and the system is in the chiral phase, with $\boldsymbol{\psi}\propto (1,i,0)$. In this 
case the system is invariant under rotations around ${\bf \Sigma}_2$ combined with a gauge transformation \cite{SuppMat}. Finally, 
when $d_2>0$ and $b_2'<0$ one has ${\bf \Sigma}_2\neq 0$, but $\boldsymbol{\psi}$ is not in the purely chiral state, but rather in 
a hybrid solution \cite{SuppMat} which has no symmetry. We name this phase TRSB 2.  

\begin{figure}[t!]
\includegraphics[width=8cm]{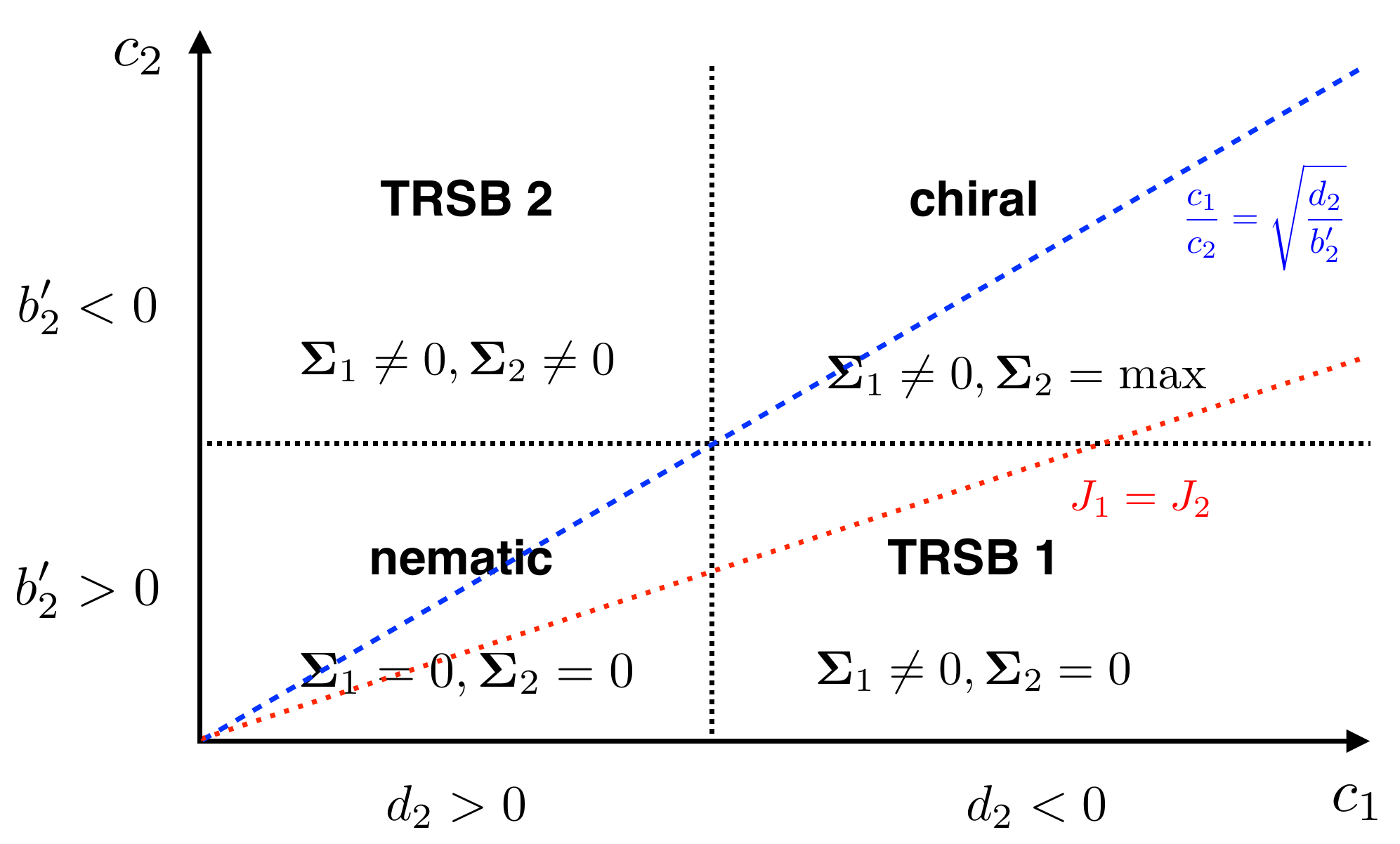}
\caption{Phase diagram of superconductivity involving the pseudo-scalar and the vector order parameters coupled to the dopants 
magnetization. The four possible phases can be obtained by properly tuning the couplings $J_1$ and $J_2$.}
\label{phasediagram}
\end{figure}

A schematic phase diagram as a function of $b_2'$ and $d_2$ is depicted in Fig. \ref{phasediagram}.  
Microscopic calculations \cite{FuPRBR2014,Venderbos2015,SuppMat} show that for an isotropic model $b_2',d_2>0$, precluding a TRSB phase. We show next how a coupling to magnetic dopants renormalizes the coefficients $b_2'$ and $d_2$ and can change their sign if the coupling is strong enough.

{\it Coupling to dopant magnetization --} The presence of random magnetic moments in the sample can be described by an average magnetization ${\bf M}$. At the Landau theory level, both ${\bf \Sigma}_1$ and ${\bf \Sigma}_2$ can couple linearly to 
${\bf M}$ \cite{WalkerPRL2002,Mineev2002,WalkerPRB2002,Mineev2004}  which is also a ${\cal T}$-odd pseudo-vector
\begin{equation}\label{Eq:3rdOrder}
F_{\chi,\psi,M}=i{\bf M}\cdot[c_1(\chi \boldsymbol{\psi}^* - \chi^*\boldsymbol{\psi})+c_2\boldsymbol{\psi}\times \boldsymbol{\psi}^*].
\end{equation}
By appropriately aligning ${\bf M}$, we see that the system may lower its energy by condensing in a TRSB phase with finite condensate 
magnetizations. 

Neglecting interactions between the magnetic moments, the full free energy at second order in ${\bf M}$ including the
superconducting order parameters reads 
\begin{equation}\label{FullGL}
F = a_3 |{\bf M}|^2 + F_\chi+F_\psi+F_{\chi,\psi}+F_{\chi,\psi,M}.
\end{equation}
Since the dopants are paramagnetic above $T_c$, we assume $a_3>0$. The mean-field solution for ${\bf M}$ can be found by 
minimizing the free energy with respect to ${\bf M}$, finding ${\bf M}=-i\frac{c_1}{2a_3}{\bf \Sigma}_1-i\frac{c_2}{2a_3}{\bf \Sigma}_2$. 
It is clear that a non-zero magnetization ${\bf M}$ arises in all TRSB phases, despite the fact that the dopants are initially paramagnetic. Substituting the mean-field value of the magnetization the free energy takes the form of Eq.~(\ref{Fchipsi}) with 
modified parameters
\begin{equation}
d_2 \to d_2-\frac{c_1^2}{4a_3},\qquad  b_2' \to b_2' -\frac{c^2_2}{4a_3}.
\end{equation}
Since the coupling to magnetic dopants renormalizes both $b_2'$ and $d_2$, with different values of $c_1$ and $c_2$ one can now span the entire phase diagram in Fig.~\ref{phasediagram}. 

{\it Meissner screening} - The presence of the magnetic dopants induces the condensation of a TRSB phase where the dopants moments are aligned with the spin magnetization of the condensate. 
The resulting total spin magnetization ${\bf M}_{\rm s}={\bf M}+i\mu({\bf \Sigma}_1+{\bf \Sigma}_2)$ acts back onto the orbital degrees of freedom and the GL free energy is \cite{Ginzburg57,Shopova}
\begin{equation}
{\cal F}=\int d{\bf r}~\left[F+\frac{{\bf B}^2}{8\pi}-{\bf B}\cdot{\bf M}_{\rm s}+F^{\rm grad}_{\chi,\boldsymbol{\psi},{\bf M}}\right].
\end{equation}
where ${\bf B}$ is the full induction field and $F^{\rm grad}$ accounts for gradient terms for the order parameters \cite{SuppMat}. 
For finite ${\bf M}_{\rm s}$ the system may develop screening supercurrents, so that ${\bf B}={\bf H}+4\pi({\bf M}_{\rm s}+{\bf M}_{\rm o})$, with ${\bf M}_{\rm o}$ the orbital magnetization due to screening currents, and ${\bf H}$ an external field.  For ${\bf H}=0$, the order parameters in the bulk can be taken to be constant, so that ${\bf B}=0$ by Meissner screening, provided that $M_{\rm s}<H_{\rm cr}$, 
with $H_{\rm cr}$ the thermodynamic critical field \cite{Ginzburg57,SuppMat}. Since $M_{\rm s}$ is linked to the mean-field value of $\chi$ and 
$\boldsymbol{\psi}$,  for $a_1\sim a_2$ the ratio $M_{\rm s}/H_{\rm cr}$ is temperature independent and it is suppressed by strong 
$b_1$ and $b_2$. At the surface of the system the cancelation between spin and orbital magnetization is not satisfied locally, due to difference in the coherence length, penetration depth, and the length scale of variation of ${\bf M}$, and a finite surface magnetization may arise, in agreement with the observations of Ref.~\cite{Qiu}.

{\it Microscopic coupling --}
The coupling Eq.~(\ref{Eq:3rdOrder}) and the resulting phase diagram is generic of a SO(3) invariant theory. 
The only symmetry allowed microscopic coupling must be written in terms of the spin pseudovectors ${\bf S}_\parallel\equiv(s_x,s_y,\sigma_xs_z)$ 
and ${\bf S}_\perp\equiv(\sigma_xs_x,\sigma_xs_y,s_z)$ \cite{SuppMat},
\begin{equation}
H_{\rm Z}=J_1{\bf M}\cdot{\bf S}_\parallel+J_2{\bf M}\cdot{\bf S}_\perp.
\end{equation}
The coefficients $c_1$ and $c_2$ can be derived microscopically from this coupling, and doing so reveals the constraint $c_1(J_1m/\mu+J_2)=2c_2(J_1+J_2m/\mu)$ \cite{SuppMat}. All phases in Fig.~\ref{phasediagram} 
can therefore be realized by properly tuning $J_1$, $J_2$, and $m/\mu$. 
In Bi$_2$Se$_3$, the SO(3) symmetry breaks down to the lattice point group $D_{3d}$ 
when anisotropy corrections are included~\cite{FuBerg}. The vector $\boldsymbol{\psi} = (\psi_x,\psi_y,\psi_z)$ 
splits into a two-component $E_{u} \sim (-\psi_y,\psi_x)$ and one-component $A_{2u}\sim \psi_z$ representations. 
The pseudoscalar $\chi$ corresponds to the $A_{1u}$ representation. A microscopic coupling between the magnetic 
moments and the physical spin ${\bf s}$ of the electrons in Bi$_2$Se$_3$ can be written in terms of a Zeeman coupling 
$H_{\rm Z}=-J(s_xM_x+s_yM_y)-J_zs_zM_z$ with $J\neq J_z$ anisotropic Zeeman coupling constants. The resulting phase 
diagram remains qualitatively very similar to the SO(3) invariant one \cite{SuppMat}.  

\begin{figure}[t!]
\includegraphics[width=8cm]{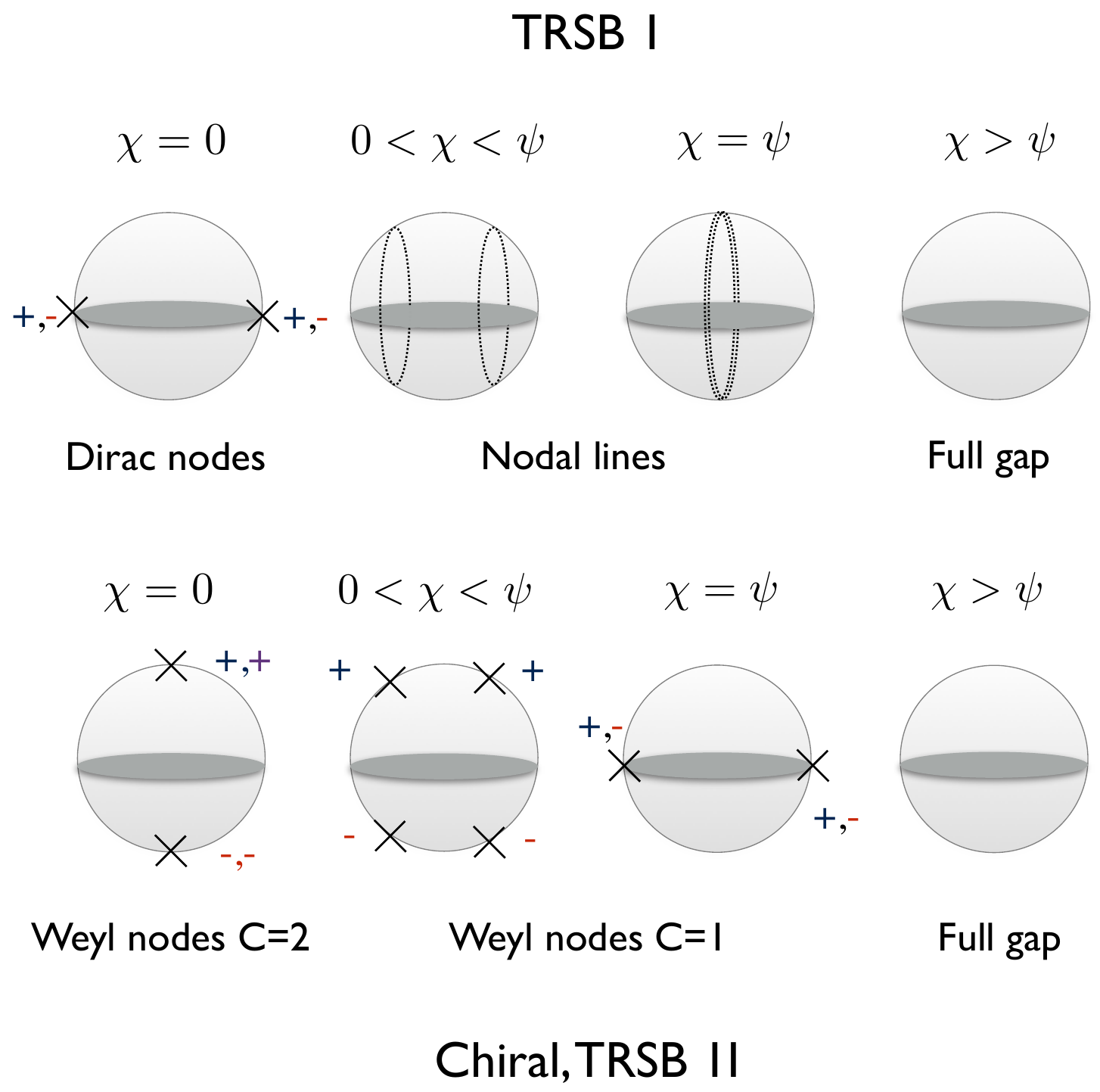}
\caption{Schematics of the gap structure on the Fermi surface: TRSB 1 has Dirac nodes that evolve in nodal line for $\chi>0$. TRSB 2 and 
the chiral phase have Weyl points with $C=2$ that split in two $C=1$ upon switching $\chi$. The phases are fully gapped for $\chi>\psi$.}
\label{Fig:Gap}
\end{figure}

{\it Gap structure --} The value of the superconducting gap on the Fermi surface for the different phases depends on the relative strength of the two order parameters. When $\chi$ dominates all phases are fully gapped, but different cases arise if $\boldsymbol{\psi}$ dominates. In the nematic case the gap has Dirac nodes along the nematic direction for $\chi=0$. These nodes can be gapped by a 
small $\chi$ or by hexagonal warping terms \cite{FuPRBR2014}, so that in general the phase is fully gapped. In the TRSB 1 
phase the order parameters may be taken as $\boldsymbol{\psi} =\psi_0 (1,0,0)$ and $\chi =\chi_0 e^{i\gamma}$ and that the Dirac nodes for $\chi=0$ 
can be shown to become circular nodal lines defined by $\sin\theta=\pm\chi_0/\psi_0$, with $\theta$ the polar angle with respect to $\boldsymbol{\Sigma}_1$. 
Nodal lines of the north and south hemisphere join for $\chi=\psi$ and become gapped for $\chi>\psi$ (see Fig.~\ref{Fig:Gap}). These nodal lines have a linear density of states (DOS) $\rho(\epsilon)\propto\epsilon$ \cite{Phillips2014}. In the chiral and the TRSB 2 phase a Weyl superconductor is realized
\cite{Meng2012,Sau2012,Yang2014,Venderbos2015}. For $\chi=0$ there are Weyl nodes of topological charge $C=\pm 2$ on the north and south 
pole along the direction of ${\bf \Sigma}_2$ \cite{Kozii2016}. For finite $\chi$ these nodes are split into two Weyl nodes of $C=1$ at a finite polar angle and in the azimuthal direction given by $\Sigma_1$ and by increasing $\chi$ they move towards the equator where they meet with the nodes from the south hemisphere and gap out for $\chi>\psi$ (see Fig.~\ref{Fig:Gap}). 
Note that while the DOS is linear in energy when $\chi=0$, $\rho_{C=2}(\epsilon)\propto\epsilon$, it becomes quadratic for finite $\chi$, 
$\rho_{C=1}(\epsilon)\propto\epsilon^2$ \cite{Fang2012}. These predictions could be confirmed by STM or specific heat measurements. 
On the surface of Weyl superconductor there are Majorana arcs of different kinds \cite{Kozii2016}, while in the gapped phases the topologically 
protected surface Andreev states associated to $\chi$ are gapped on the surfaces orthogonal to ${\bf \Sigma}_1$.

{\it Discussion and conclusions--} 
The features of the TRSB2 phase predicted in this work are consistent with all the observations made in recent experiments with Nb$_x$Bi$_2$Se$_3$: the breaking of rotation~\cite{Asaba} and time-reversal symmetry~\cite{Qiu} and the presence of point nodes~\cite{Smylie}. These conclusions remain valid also if the scalar and vector representations are split due lattice symmetries. In this case, the lattice will naturally pin the direction of $\Sigma_2$ to the $c$ axis, while $\Sigma_1$ will lay in-plane, pointing in a high-symmetry direction. This is enough to reproduce the twofold pattern observed in torque magnetometry. Our work makes the additional prediction that the magnetization, which can only be observed in the surface due to Meissner screening, must have both in-plane and out-of-plane components. The TRSB2 phase also features linear nodes in the bulk with Chern number $C=1$, consistent with the scaling of the penetration depth. This is in contrast with the TRI nematic candidate state, which was argued to be fully gapped in the presence of trigonal warping \cite{FuPRBR2014}. Our work further predicts the positions of the nodes to lie in the direction of $\Sigma_1$, a prediction that could be tested, for example, with the nodal spectroscopy techniques proposed in Refs. \cite{Valls1,Valls2,Valls3}. Finally, our work also provides a general framework to address current and future experiments with doped Dirac materials, emphasizing the importance of mixed symmetry states and coexistence of order parameters.

{\it Note --} During the preparation of this manuscript, we became aware of Ref. \cite{Yuan}, where magnetic Nb dopants are also considered as the mechanism that stabilizes chiral superconductivity. This work does not provide a mechanism for the vector channel to compete with the pseudoscalar, and no mixed symmetry phases are considered. The chiral state proposed in Ref. \cite{Yuan} respects $C_3$ rotation symmetry, in contrast with Ref.~\cite{Asaba}. The issue of Meissner screening is also not addressed.

{\it Acknowledgements -- } The authors acknowledge useful discussions with Irina Grigorieva. The authors acknowledge
funding from the European Union's Seventh Framework Programme (FP7/2007-2013) through the ERC Advanced Grant 
NOVGRAPHENE through grant agreement Nr. 290846 (L. C., F. J. and F. G.), from the Marie Curie Programme under EC Grant agreement No. 705968 (F. J.) and from the European Commission under the Graphene Flagship, contract CNECTICT-604391 (F. G.).

\bibliography{Bib-TRSB_nourl}

\begin{thebibliography}{49}
\expandafter\ifx\csname natexlab\endcsname\relax\def\natexlab#1{#1}\fi
\expandafter\ifx\csname bibnamefont\endcsname\relax
  \def\bibnamefont#1{#1}\fi
\expandafter\ifx\csname bibfnamefont\endcsname\relax
  \def\bibfnamefont#1{#1}\fi
\expandafter\ifx\csname citenamefont\endcsname\relax
  \def\citenamefont#1{#1}\fi
\expandafter\ifx\csname url\endcsname\relax
  \def\url#1{\texttt{#1}}\fi
\expandafter\ifx\csname urlprefix\endcsname\relax\def\urlprefix{URL }\fi
\providecommand{\bibinfo}[2]{#2}
\providecommand{\eprint}[2][]{\url{#2}}

\bibitem[{\citenamefont{Sigrist and Ueda}(1991)}]{SigristUeda}
\bibinfo{author}{\bibfnamefont{M.}~\bibnamefont{Sigrist}} \bibnamefont{and}
  \bibinfo{author}{\bibfnamefont{K.}~\bibnamefont{Ueda}},
  \bibinfo{journal}{Rev. Mod. Phys.} \textbf{\bibinfo{volume}{63}},
  \bibinfo{pages}{239} (\bibinfo{year}{1991}).

\bibitem[{\citenamefont{Maeno et~al.}(1994)\citenamefont{Maeno, Hashimoto,
  Yoshida, Nishizaki, Fujita, Bednorz, and Lichtenberg}}]{Maeno}
\bibinfo{author}{\bibfnamefont{Y.}~\bibnamefont{Maeno}},
  \bibinfo{author}{\bibfnamefont{H.}~\bibnamefont{Hashimoto}},
  \bibinfo{author}{\bibfnamefont{K.}~\bibnamefont{Yoshida}},
  \bibinfo{author}{\bibfnamefont{S.}~\bibnamefont{Nishizaki}},
  \bibinfo{author}{\bibfnamefont{T.}~\bibnamefont{Fujita}},
  \bibinfo{author}{\bibfnamefont{J.~G.} \bibnamefont{Bednorz}},
  \bibnamefont{and}
  \bibinfo{author}{\bibfnamefont{F.}~\bibnamefont{Lichtenberg}},
  \bibinfo{journal}{Nature} \textbf{\bibinfo{volume}{372}},
  \bibinfo{pages}{532} (\bibinfo{year}{1994}).

\bibitem[{\citenamefont{Luke et~al.}(1993)\citenamefont{Luke, Keren, Le, Wu,
  Uemura, Bonn, Taillefer, and Garrett}}]{Luke1993}
\bibinfo{author}{\bibfnamefont{G.~M.} \bibnamefont{Luke}},
  \bibinfo{author}{\bibfnamefont{A.}~\bibnamefont{Keren}},
  \bibinfo{author}{\bibfnamefont{L.~P.} \bibnamefont{Le}},
  \bibinfo{author}{\bibfnamefont{W.~D.} \bibnamefont{Wu}},
  \bibinfo{author}{\bibfnamefont{Y.~J.} \bibnamefont{Uemura}},
  \bibinfo{author}{\bibfnamefont{D.~A.} \bibnamefont{Bonn}},
  \bibinfo{author}{\bibfnamefont{L.}~\bibnamefont{Taillefer}},
  \bibnamefont{and} \bibinfo{author}{\bibfnamefont{J.~D.}
  \bibnamefont{Garrett}}, \bibinfo{journal}{Phys. Rev. Lett.}
  \textbf{\bibinfo{volume}{71}}, \bibinfo{pages}{1466} (\bibinfo{year}{1993}).

\bibitem[{\citenamefont{Luke et~al.}(1998)\citenamefont{Luke, Fudamoto, Kojima,
  Larkin, Merrin, Nachumi, Uemura, Maeno, Mao, Mori et~al.}}]{Luke1998}
\bibinfo{author}{\bibfnamefont{G.~M.} \bibnamefont{Luke}},
  \bibinfo{author}{\bibfnamefont{Y.}~\bibnamefont{Fudamoto}},
  \bibinfo{author}{\bibfnamefont{K.~M.} \bibnamefont{Kojima}},
  \bibinfo{author}{\bibfnamefont{M.~I.} \bibnamefont{Larkin}},
  \bibinfo{author}{\bibfnamefont{J.}~\bibnamefont{Merrin}},
  \bibinfo{author}{\bibfnamefont{B.}~\bibnamefont{Nachumi}},
  \bibinfo{author}{\bibfnamefont{Y.~J.} \bibnamefont{Uemura}},
  \bibinfo{author}{\bibfnamefont{Y.}~\bibnamefont{Maeno}},
  \bibinfo{author}{\bibfnamefont{Z.~Q.} \bibnamefont{Mao}},
  \bibinfo{author}{\bibfnamefont{Y.}~\bibnamefont{Mori}}, \bibnamefont{et~al.},
  \bibinfo{journal}{Nature} \textbf{\bibinfo{volume}{394}},
  \bibinfo{pages}{558} (\bibinfo{year}{1998}).

\bibitem[{\citenamefont{Kapitulnik et~al.}(2009)\citenamefont{Kapitulnik, Xia,
  Schemm, and Palevski}}]{Kapitulnik}
\bibinfo{author}{\bibfnamefont{A.}~\bibnamefont{Kapitulnik}},
  \bibinfo{author}{\bibfnamefont{J.}~\bibnamefont{Xia}},
  \bibinfo{author}{\bibfnamefont{E.}~\bibnamefont{Schemm}}, \bibnamefont{and}
  \bibinfo{author}{\bibfnamefont{A.}~\bibnamefont{Palevski}},
  \bibinfo{journal}{New Journal of Physics} \textbf{\bibinfo{volume}{11}},
  \bibinfo{pages}{055060} (\bibinfo{year}{2009}).

\bibitem[{\citenamefont{Nayak et~al.}(2008)\citenamefont{Nayak, Simon, Stern,
  Freedman, and Das~Sarma}}]{Nayak}
\bibinfo{author}{\bibfnamefont{C.}~\bibnamefont{Nayak}},
  \bibinfo{author}{\bibfnamefont{S.~H.} \bibnamefont{Simon}},
  \bibinfo{author}{\bibfnamefont{A.}~\bibnamefont{Stern}},
  \bibinfo{author}{\bibfnamefont{M.}~\bibnamefont{Freedman}}, \bibnamefont{and}
  \bibinfo{author}{\bibfnamefont{S.}~\bibnamefont{Das~Sarma}},
  \bibinfo{journal}{Rev. Mod. Phys.} \textbf{\bibinfo{volume}{80}},
  \bibinfo{pages}{1083} (\bibinfo{year}{2008}).

\bibitem[{\citenamefont{Qi and Zhang}(2011)}]{Qi}
\bibinfo{author}{\bibfnamefont{X.-L.} \bibnamefont{Qi}} \bibnamefont{and}
  \bibinfo{author}{\bibfnamefont{S.-C.} \bibnamefont{Zhang}},
  \bibinfo{journal}{Rev. Mod. Phys.} \textbf{\bibinfo{volume}{83}},
  \bibinfo{pages}{1057} (\bibinfo{year}{2011}).

\bibitem[{\citenamefont{Meng and Balents}(2012)}]{Meng2012}
\bibinfo{author}{\bibfnamefont{T.}~\bibnamefont{Meng}} \bibnamefont{and}
  \bibinfo{author}{\bibfnamefont{L.}~\bibnamefont{Balents}},
  \bibinfo{journal}{Phys. Rev. B} \textbf{\bibinfo{volume}{86}},
  \bibinfo{pages}{054504} (\bibinfo{year}{2012}).

\bibitem[{\citenamefont{Sau and Tewari}(2012)}]{Sau2012}
\bibinfo{author}{\bibfnamefont{J.~D.} \bibnamefont{Sau}} \bibnamefont{and}
  \bibinfo{author}{\bibfnamefont{S.}~\bibnamefont{Tewari}},
  \bibinfo{journal}{Phys. Rev. B} \textbf{\bibinfo{volume}{86}},
  \bibinfo{pages}{104509} (\bibinfo{year}{2012}).

\bibitem[{\citenamefont{Yang et~al.}(2014)\citenamefont{Yang, Pan, and
  Zhang}}]{Yang2014}
\bibinfo{author}{\bibfnamefont{S.~A.} \bibnamefont{Yang}},
  \bibinfo{author}{\bibfnamefont{H.}~\bibnamefont{Pan}}, \bibnamefont{and}
  \bibinfo{author}{\bibfnamefont{F.}~\bibnamefont{Zhang}},
  \bibinfo{journal}{Phys. Rev. Lett.} \textbf{\bibinfo{volume}{113}},
  \bibinfo{pages}{046401} (\bibinfo{year}{2014}).

\bibitem[{\citenamefont{Zhang et~al.}(2009)\citenamefont{Zhang, Liu, Qi, Dai,
  Fang, and Zhang}}]{Zhang-NP2009}
\bibinfo{author}{\bibfnamefont{H.}~\bibnamefont{Zhang}},
  \bibinfo{author}{\bibfnamefont{C.-X.} \bibnamefont{Liu}},
  \bibinfo{author}{\bibfnamefont{X.-L.} \bibnamefont{Qi}},
  \bibinfo{author}{\bibfnamefont{X.}~\bibnamefont{Dai}},
  \bibinfo{author}{\bibfnamefont{Z.}~\bibnamefont{Fang}}, \bibnamefont{and}
  \bibinfo{author}{\bibfnamefont{S.-C.} \bibnamefont{Zhang}},
  \bibinfo{journal}{Nat Phys} \textbf{\bibinfo{volume}{5}},
  \bibinfo{pages}{438} (\bibinfo{year}{2009}).

\bibitem[{\citenamefont{Hasan and Kane}(2010)}]{Hasan2010}
\bibinfo{author}{\bibfnamefont{M.~Z.} \bibnamefont{Hasan}} \bibnamefont{and}
  \bibinfo{author}{\bibfnamefont{C.~L.} \bibnamefont{Kane}},
  \bibinfo{journal}{Rev. Mod. Phys.} \textbf{\bibinfo{volume}{82}},
  \bibinfo{pages}{3045} (\bibinfo{year}{2010}).

\bibitem[{\citenamefont{Fu and Berg}(2010)}]{FuBerg}
\bibinfo{author}{\bibfnamefont{L.}~\bibnamefont{Fu}} \bibnamefont{and}
  \bibinfo{author}{\bibfnamefont{E.}~\bibnamefont{Berg}},
  \bibinfo{journal}{Phys. Rev. Lett.} \textbf{\bibinfo{volume}{105}},
  \bibinfo{pages}{097001} (\bibinfo{year}{2010}).

\bibitem[{\citenamefont{Hor et~al.}(2010)\citenamefont{Hor, Williams,
  Checkelsky, Roushan, Seo, Xu, Zandbergen, Yazdani, Ong, and
  Cava}}]{Hor-PRL2010}
\bibinfo{author}{\bibfnamefont{Y.~S.} \bibnamefont{Hor}},
  \bibinfo{author}{\bibfnamefont{A.~J.} \bibnamefont{Williams}},
  \bibinfo{author}{\bibfnamefont{J.~G.} \bibnamefont{Checkelsky}},
  \bibinfo{author}{\bibfnamefont{P.}~\bibnamefont{Roushan}},
  \bibinfo{author}{\bibfnamefont{J.}~\bibnamefont{Seo}},
  \bibinfo{author}{\bibfnamefont{Q.}~\bibnamefont{Xu}},
  \bibinfo{author}{\bibfnamefont{H.~W.} \bibnamefont{Zandbergen}},
  \bibinfo{author}{\bibfnamefont{A.}~\bibnamefont{Yazdani}},
  \bibinfo{author}{\bibfnamefont{N.~P.} \bibnamefont{Ong}}, \bibnamefont{and}
  \bibinfo{author}{\bibfnamefont{R.~J.} \bibnamefont{Cava}},
  \bibinfo{journal}{Phys. Rev. Lett.} \textbf{\bibinfo{volume}{104}},
  \bibinfo{pages}{057001} (\bibinfo{year}{2010}).

\bibitem[{\citenamefont{Wray et~al.}(2010)\citenamefont{Wray, Xu, Xia, Hor,
  Qian, Fedorov, Lin, Bansil, Cava, and Hasan}}]{Wray-NP2010}
\bibinfo{author}{\bibfnamefont{L.~A.} \bibnamefont{Wray}},
  \bibinfo{author}{\bibfnamefont{S.-Y.} \bibnamefont{Xu}},
  \bibinfo{author}{\bibfnamefont{Y.}~\bibnamefont{Xia}},
  \bibinfo{author}{\bibfnamefont{Y.~S.} \bibnamefont{Hor}},
  \bibinfo{author}{\bibfnamefont{D.}~\bibnamefont{Qian}},
  \bibinfo{author}{\bibfnamefont{A.~V.} \bibnamefont{Fedorov}},
  \bibinfo{author}{\bibfnamefont{H.}~\bibnamefont{Lin}},
  \bibinfo{author}{\bibfnamefont{A.}~\bibnamefont{Bansil}},
  \bibinfo{author}{\bibfnamefont{R.~J.} \bibnamefont{Cava}}, \bibnamefont{and}
  \bibinfo{author}{\bibfnamefont{M.~Z.} \bibnamefont{Hasan}},
  \bibinfo{journal}{Nat Phys} \textbf{\bibinfo{volume}{6}},
  \bibinfo{pages}{855} (\bibinfo{year}{2010}).

\bibitem[{\citenamefont{Kriener et~al.}(2011)\citenamefont{Kriener, Segawa,
  Ren, Sasaki, and Ando}}]{Kriener-PRL2011}
\bibinfo{author}{\bibfnamefont{M.}~\bibnamefont{Kriener}},
  \bibinfo{author}{\bibfnamefont{K.}~\bibnamefont{Segawa}},
  \bibinfo{author}{\bibfnamefont{Z.}~\bibnamefont{Ren}},
  \bibinfo{author}{\bibfnamefont{S.}~\bibnamefont{Sasaki}}, \bibnamefont{and}
  \bibinfo{author}{\bibfnamefont{Y.}~\bibnamefont{Ando}},
  \bibinfo{journal}{Phys. Rev. Lett.} \textbf{\bibinfo{volume}{106}},
  \bibinfo{pages}{127004} (\bibinfo{year}{2011}).

\bibitem[{\citenamefont{Sasaki et~al.}(2011)\citenamefont{Sasaki, Kriener,
  Segawa, Yada, Tanaka, Sato, and Ando}}]{Sasaki-PRL2011}
\bibinfo{author}{\bibfnamefont{S.}~\bibnamefont{Sasaki}},
  \bibinfo{author}{\bibfnamefont{M.}~\bibnamefont{Kriener}},
  \bibinfo{author}{\bibfnamefont{K.}~\bibnamefont{Segawa}},
  \bibinfo{author}{\bibfnamefont{K.}~\bibnamefont{Yada}},
  \bibinfo{author}{\bibfnamefont{Y.}~\bibnamefont{Tanaka}},
  \bibinfo{author}{\bibfnamefont{M.}~\bibnamefont{Sato}}, \bibnamefont{and}
  \bibinfo{author}{\bibfnamefont{Y.}~\bibnamefont{Ando}},
  \bibinfo{journal}{Phys. Rev. Lett.} \textbf{\bibinfo{volume}{107}},
  \bibinfo{pages}{217001} (\bibinfo{year}{2011}).

\bibitem[{\citenamefont{Levy et~al.}(2013)\citenamefont{Levy, Zhang, Ha,
  Sharifi, Talin, Kuk, and Stroscio}}]{Levy-PRL2013}
\bibinfo{author}{\bibfnamefont{N.}~\bibnamefont{Levy}},
  \bibinfo{author}{\bibfnamefont{T.}~\bibnamefont{Zhang}},
  \bibinfo{author}{\bibfnamefont{J.}~\bibnamefont{Ha}},
  \bibinfo{author}{\bibfnamefont{F.}~\bibnamefont{Sharifi}},
  \bibinfo{author}{\bibfnamefont{A.~A.} \bibnamefont{Talin}},
  \bibinfo{author}{\bibfnamefont{Y.}~\bibnamefont{Kuk}}, \bibnamefont{and}
  \bibinfo{author}{\bibfnamefont{J.~A.} \bibnamefont{Stroscio}},
  \bibinfo{journal}{Phys. Rev. Lett.} \textbf{\bibinfo{volume}{110}},
  \bibinfo{pages}{117001} (\bibinfo{year}{2013}).

\bibitem[{\citenamefont{Peng et~al.}(2013)\citenamefont{Peng, De, Lv, Wei, and
  Chu}}]{Peng-PRB2013}
\bibinfo{author}{\bibfnamefont{H.}~\bibnamefont{Peng}},
  \bibinfo{author}{\bibfnamefont{D.}~\bibnamefont{De}},
  \bibinfo{author}{\bibfnamefont{B.}~\bibnamefont{Lv}},
  \bibinfo{author}{\bibfnamefont{F.}~\bibnamefont{Wei}}, \bibnamefont{and}
  \bibinfo{author}{\bibfnamefont{C.-W.} \bibnamefont{Chu}},
  \bibinfo{journal}{Phys. Rev. B} \textbf{\bibinfo{volume}{88}},
  \bibinfo{pages}{024515} (\bibinfo{year}{2013}).

\bibitem[{\citenamefont{Matano et~al.}(2016)\citenamefont{Matano, Kriener,
  Segawa, Ando, and Zheng}}]{Matano}
\bibinfo{author}{\bibfnamefont{K.}~\bibnamefont{Matano}},
  \bibinfo{author}{\bibfnamefont{M.}~\bibnamefont{Kriener}},
  \bibinfo{author}{\bibfnamefont{K.}~\bibnamefont{Segawa}},
  \bibinfo{author}{\bibfnamefont{Y.}~\bibnamefont{Ando}}, \bibnamefont{and}
  \bibinfo{author}{\bibfnamefont{G.-q.} \bibnamefont{Zheng}},
  \bibinfo{journal}{Nat Phys} \textbf{\bibinfo{volume}{12}},
  \bibinfo{pages}{852} (\bibinfo{year}{2016}).

\bibitem[{\citenamefont{Fu}(2014)}]{FuPRBR2014}
\bibinfo{author}{\bibfnamefont{L.}~\bibnamefont{Fu}}, \bibinfo{journal}{Phys.
  Rev. B} \textbf{\bibinfo{volume}{90}}, \bibinfo{pages}{100509}
  (\bibinfo{year}{2014}).

\bibitem[{\citenamefont{Venderbos
  et~al.}(2016{\natexlab{a}})\citenamefont{Venderbos, Kozii, and
  Fu}}]{Venderbos2016}
\bibinfo{author}{\bibfnamefont{J.~W.~F.} \bibnamefont{Venderbos}},
  \bibinfo{author}{\bibfnamefont{V.}~\bibnamefont{Kozii}}, \bibnamefont{and}
  \bibinfo{author}{\bibfnamefont{L.}~\bibnamefont{Fu}}, \bibinfo{journal}{Phys.
  Rev. B} \textbf{\bibinfo{volume}{94}}, \bibinfo{pages}{094522}
  (\bibinfo{year}{2016}{\natexlab{a}}).

\bibitem[{\citenamefont{Shruti et~al.}(2015)\citenamefont{Shruti, Maurya, Neha,
  Srivastava, and Patnaik}}]{Shruti}
\bibinfo{author}{\bibnamefont{Shruti}}, \bibinfo{author}{\bibfnamefont{V.~K.}
  \bibnamefont{Maurya}},
  \bibinfo{author}{\bibfnamefont{P.}~\bibnamefont{Neha}},
  \bibinfo{author}{\bibfnamefont{P.}~\bibnamefont{Srivastava}},
  \bibnamefont{and} \bibinfo{author}{\bibfnamefont{S.}~\bibnamefont{Patnaik}},
  \bibinfo{journal}{Phys. Rev. B} \textbf{\bibinfo{volume}{92}},
  \bibinfo{pages}{020506} (\bibinfo{year}{2015}).

\bibitem[{\citenamefont{Liu et~al.}(2015)\citenamefont{Liu, Yao, Shao, Zuo, Pi,
  Tan, Zhang, and Zhang}}]{Liu}
\bibinfo{author}{\bibfnamefont{Z.}~\bibnamefont{Liu}},
  \bibinfo{author}{\bibfnamefont{X.}~\bibnamefont{Yao}},
  \bibinfo{author}{\bibfnamefont{J.}~\bibnamefont{Shao}},
  \bibinfo{author}{\bibfnamefont{M.}~\bibnamefont{Zuo}},
  \bibinfo{author}{\bibfnamefont{L.}~\bibnamefont{Pi}},
  \bibinfo{author}{\bibfnamefont{S.}~\bibnamefont{Tan}},
  \bibinfo{author}{\bibfnamefont{C.}~\bibnamefont{Zhang}}, \bibnamefont{and}
  \bibinfo{author}{\bibfnamefont{Y.}~\bibnamefont{Zhang}},
  \bibinfo{journal}{Journal of the American Chemical Society}
  \textbf{\bibinfo{volume}{137}}, \bibinfo{pages}{10512}
  (\bibinfo{year}{2015}).

\bibitem[{\citenamefont{Wang et~al.}(2016)\citenamefont{Wang, Taskin,
  Fr{\"o}lich, Braden, and Ando}}]{Wang}
\bibinfo{author}{\bibfnamefont{Z.}~\bibnamefont{Wang}},
  \bibinfo{author}{\bibfnamefont{A.~A.} \bibnamefont{Taskin}},
  \bibinfo{author}{\bibfnamefont{T.}~\bibnamefont{Fr{\"o}lich}},
  \bibinfo{author}{\bibfnamefont{M.}~\bibnamefont{Braden}}, \bibnamefont{and}
  \bibinfo{author}{\bibfnamefont{Y.}~\bibnamefont{Ando}},
  \bibinfo{journal}{Chemistry of Materials} \textbf{\bibinfo{volume}{28}},
  \bibinfo{pages}{779} (\bibinfo{year}{2016}).

\bibitem[{\citenamefont{{Qiu} et~al.}(2015)\citenamefont{{Qiu}, {Nocona
  Sanders}, {Dai}, {Medvedeva}, {Wu}, {Ghaemi}, {Vojta}, and {San Hor}}}]{Qiu}
\bibinfo{author}{\bibfnamefont{Y.}~\bibnamefont{{Qiu}}},
  \bibinfo{author}{\bibfnamefont{K.}~\bibnamefont{{Nocona Sanders}}},
  \bibinfo{author}{\bibfnamefont{J.}~\bibnamefont{{Dai}}},
  \bibinfo{author}{\bibfnamefont{J.~E.} \bibnamefont{{Medvedeva}}},
  \bibinfo{author}{\bibfnamefont{W.}~\bibnamefont{{Wu}}},
  \bibinfo{author}{\bibfnamefont{P.}~\bibnamefont{{Ghaemi}}},
  \bibinfo{author}{\bibfnamefont{T.}~\bibnamefont{{Vojta}}}, \bibnamefont{and}
  \bibinfo{author}{\bibfnamefont{Y.}~\bibnamefont{{San Hor}}},
  \bibinfo{journal}{ArXiv e-prints}  (\bibinfo{year}{2015}),
  \eprint{1512.03519}.

\bibitem[{\citenamefont{Asaba et~al.}(2017)\citenamefont{Asaba, Lawson,
  Tinsman, Chen, Corbae, Li, Qiu, Hor, Fu, and Li}}]{Asaba}
\bibinfo{author}{\bibfnamefont{T.}~\bibnamefont{Asaba}},
  \bibinfo{author}{\bibfnamefont{B.~J.} \bibnamefont{Lawson}},
  \bibinfo{author}{\bibfnamefont{C.}~\bibnamefont{Tinsman}},
  \bibinfo{author}{\bibfnamefont{L.}~\bibnamefont{Chen}},
  \bibinfo{author}{\bibfnamefont{P.}~\bibnamefont{Corbae}},
  \bibinfo{author}{\bibfnamefont{G.}~\bibnamefont{Li}},
  \bibinfo{author}{\bibfnamefont{Y.}~\bibnamefont{Qiu}},
  \bibinfo{author}{\bibfnamefont{Y.~S.} \bibnamefont{Hor}},
  \bibinfo{author}{\bibfnamefont{L.}~\bibnamefont{Fu}}, \bibnamefont{and}
  \bibinfo{author}{\bibfnamefont{L.}~\bibnamefont{Li}}, \bibinfo{journal}{Phys.
  Rev. X} \textbf{\bibinfo{volume}{7}}, \bibinfo{pages}{011009}
  (\bibinfo{year}{2017}).

\bibitem[{\citenamefont{Smylie et~al.}(2016)\citenamefont{Smylie, Claus, Welp,
  Kwok, Qiu, Hor, and Snezhko}}]{Smylie}
\bibinfo{author}{\bibfnamefont{M.~P.} \bibnamefont{Smylie}},
  \bibinfo{author}{\bibfnamefont{H.}~\bibnamefont{Claus}},
  \bibinfo{author}{\bibfnamefont{U.}~\bibnamefont{Welp}},
  \bibinfo{author}{\bibfnamefont{W.-K.} \bibnamefont{Kwok}},
  \bibinfo{author}{\bibfnamefont{Y.}~\bibnamefont{Qiu}},
  \bibinfo{author}{\bibfnamefont{Y.~S.} \bibnamefont{Hor}}, \bibnamefont{and}
  \bibinfo{author}{\bibfnamefont{A.}~\bibnamefont{Snezhko}},
  \bibinfo{journal}{Phys. Rev. B} \textbf{\bibinfo{volume}{94}},
  \bibinfo{pages}{180510} (\bibinfo{year}{2016}).

\bibitem[{\citenamefont{Venderbos
  et~al.}(2016{\natexlab{b}})\citenamefont{Venderbos, Kozii, and
  Fu}}]{Venderbos2015}
\bibinfo{author}{\bibfnamefont{J.~W.~F.} \bibnamefont{Venderbos}},
  \bibinfo{author}{\bibfnamefont{V.}~\bibnamefont{Kozii}}, \bibnamefont{and}
  \bibinfo{author}{\bibfnamefont{L.}~\bibnamefont{Fu}}, \bibinfo{journal}{Phys.
  Rev. B} \textbf{\bibinfo{volume}{94}}, \bibinfo{pages}{180504}
  (\bibinfo{year}{2016}{\natexlab{b}}).

\bibitem[{\citenamefont{Kotliar}(1988)}]{sid1}
\bibinfo{author}{\bibfnamefont{G.}~\bibnamefont{Kotliar}},
  \bibinfo{journal}{Phys. Rev. B} \textbf{\bibinfo{volume}{37}},
  \bibinfo{pages}{3664} (\bibinfo{year}{1988}).

\bibitem[{\citenamefont{Musaelian et~al.}(1996)\citenamefont{Musaelian,
  Betouras, Chubukov, and Joynt}}]{sid2}
\bibinfo{author}{\bibfnamefont{K.~A.} \bibnamefont{Musaelian}},
  \bibinfo{author}{\bibfnamefont{J.}~\bibnamefont{Betouras}},
  \bibinfo{author}{\bibfnamefont{A.~V.} \bibnamefont{Chubukov}},
  \bibnamefont{and} \bibinfo{author}{\bibfnamefont{R.}~\bibnamefont{Joynt}},
  \bibinfo{journal}{Phys. Rev. B} \textbf{\bibinfo{volume}{53}},
  \bibinfo{pages}{3598} (\bibinfo{year}{1996}).

\bibitem[{\citenamefont{Lee et~al.}(2009)\citenamefont{Lee, Zhang, and
  Wu}}]{sid3}
\bibinfo{author}{\bibfnamefont{W.-C.} \bibnamefont{Lee}},
  \bibinfo{author}{\bibfnamefont{S.-C.} \bibnamefont{Zhang}}, \bibnamefont{and}
  \bibinfo{author}{\bibfnamefont{C.}~\bibnamefont{Wu}}, \bibinfo{journal}{Phys.
  Rev. Lett.} \textbf{\bibinfo{volume}{102}}, \bibinfo{pages}{217002}
  (\bibinfo{year}{2009}).

\bibitem[{\citenamefont{Walker and Samokhin}(2002)}]{WalkerPRL2002}
\bibinfo{author}{\bibfnamefont{M.~B.} \bibnamefont{Walker}} \bibnamefont{and}
  \bibinfo{author}{\bibfnamefont{K.~V.} \bibnamefont{Samokhin}},
  \bibinfo{journal}{Phys. Rev. Lett.} \textbf{\bibinfo{volume}{88}},
  \bibinfo{pages}{207001} (\bibinfo{year}{2002}).

\bibitem[{\citenamefont{Mineev}(2002)}]{Mineev2002}
\bibinfo{author}{\bibfnamefont{V.~P.} \bibnamefont{Mineev}},
  \bibinfo{journal}{Phys. Rev. B} \textbf{\bibinfo{volume}{66}},
  \bibinfo{pages}{134504} (\bibinfo{year}{2002}).

\bibitem[{\citenamefont{Samokhin and Walker}(2002)}]{WalkerPRB2002}
\bibinfo{author}{\bibfnamefont{K.~V.} \bibnamefont{Samokhin}} \bibnamefont{and}
  \bibinfo{author}{\bibfnamefont{M.~B.} \bibnamefont{Walker}},
  \bibinfo{journal}{Phys. Rev. B} \textbf{\bibinfo{volume}{66}},
  \bibinfo{pages}{174501} (\bibinfo{year}{2002}).

\bibitem[{\citenamefont{MINEEV}(2004)}]{Mineev2004}
\bibinfo{author}{\bibfnamefont{V.~P.} \bibnamefont{MINEEV}},
  \bibinfo{journal}{International Journal of Modern Physics B}
  \textbf{\bibinfo{volume}{18}}, \bibinfo{pages}{2963} (\bibinfo{year}{2004}).

\bibitem[{\citenamefont{Hashimoto et~al.}(2016)\citenamefont{Hashimoto,
  Kobayashi, Tanaka, and Sato}}]{Hashimoto}
\bibinfo{author}{\bibfnamefont{T.}~\bibnamefont{Hashimoto}},
  \bibinfo{author}{\bibfnamefont{S.}~\bibnamefont{Kobayashi}},
  \bibinfo{author}{\bibfnamefont{Y.}~\bibnamefont{Tanaka}}, \bibnamefont{and}
  \bibinfo{author}{\bibfnamefont{M.}~\bibnamefont{Sato}},
  \bibinfo{journal}{Phys. Rev. B} \textbf{\bibinfo{volume}{94}},
  \bibinfo{pages}{014510} (\bibinfo{year}{2016}).

\bibitem[{Sup()}]{SuppMat}
\bibinfo{note}{See Supplementary Material for details on the Dirac matrices,
  the microscopic theory for Bi$_2$Se$_3$, the minimization of Landau free
  energies, and the gap structure on the Fermi surface.}

\bibitem[{\citenamefont{Ueda and Rice}(1985)}]{Ueda}
\bibinfo{author}{\bibfnamefont{K.}~\bibnamefont{Ueda}} \bibnamefont{and}
  \bibinfo{author}{\bibfnamefont{T.~M.} \bibnamefont{Rice}},
  \bibinfo{journal}{Phys. Rev. B} \textbf{\bibinfo{volume}{31}},
  \bibinfo{pages}{7114} (\bibinfo{year}{1985}).

\bibitem[{\citenamefont{Knigavko and Rosenstein}(1999)}]{Knigavko}
\bibinfo{author}{\bibfnamefont{A.}~\bibnamefont{Knigavko}} \bibnamefont{and}
  \bibinfo{author}{\bibfnamefont{B.}~\bibnamefont{Rosenstein}},
  \bibinfo{journal}{Phys. Rev. Lett.} \textbf{\bibinfo{volume}{82}},
  \bibinfo{pages}{1261} (\bibinfo{year}{1999}).

\bibitem[{\citenamefont{Ginzburg}(1957)}]{Ginzburg57}
\bibinfo{author}{\bibfnamefont{V.~L.} \bibnamefont{Ginzburg}},
  \bibinfo{journal}{JETP} \textbf{\bibinfo{volume}{4}}, \bibinfo{pages}{153}
  (\bibinfo{year}{1957}).

\bibitem[{\citenamefont{Shopova and Uzunov}(2005)}]{Shopova}
\bibinfo{author}{\bibfnamefont{D.~V.} \bibnamefont{Shopova}} \bibnamefont{and}
  \bibinfo{author}{\bibfnamefont{D.~I.} \bibnamefont{Uzunov}},
  \bibinfo{journal}{Phys. Rev. B} \textbf{\bibinfo{volume}{72}},
  \bibinfo{pages}{024531} (\bibinfo{year}{2005}).

\bibitem[{\citenamefont{Phillips and Aji}(2014)}]{Phillips2014}
\bibinfo{author}{\bibfnamefont{M.}~\bibnamefont{Phillips}} \bibnamefont{and}
  \bibinfo{author}{\bibfnamefont{V.}~\bibnamefont{Aji}},
  \bibinfo{journal}{Phys. Rev. B} \textbf{\bibinfo{volume}{90}},
  \bibinfo{pages}{115111} (\bibinfo{year}{2014}).

\bibitem[{\citenamefont{Kozii et~al.}(2016)\citenamefont{Kozii, Venderbos, and
  Fu}}]{Kozii2016}
\bibinfo{author}{\bibfnamefont{V.}~\bibnamefont{Kozii}},
  \bibinfo{author}{\bibfnamefont{J.~W.~F.} \bibnamefont{Venderbos}},
  \bibnamefont{and} \bibinfo{author}{\bibfnamefont{L.}~\bibnamefont{Fu}},
  \bibinfo{journal}{Science Advances} \textbf{\bibinfo{volume}{2}}
  (\bibinfo{year}{2016}).

\bibitem[{\citenamefont{Fang et~al.}(2012)\citenamefont{Fang, Gilbert, Dai, and
  Bernevig}}]{Fang2012}
\bibinfo{author}{\bibfnamefont{C.}~\bibnamefont{Fang}},
  \bibinfo{author}{\bibfnamefont{M.~J.} \bibnamefont{Gilbert}},
  \bibinfo{author}{\bibfnamefont{X.}~\bibnamefont{Dai}}, \bibnamefont{and}
  \bibinfo{author}{\bibfnamefont{B.~A.} \bibnamefont{Bernevig}},
  \bibinfo{journal}{Phys. Rev. Lett.} \textbf{\bibinfo{volume}{108}},
  \bibinfo{pages}{266802} (\bibinfo{year}{2012}).

\bibitem[{\citenamefont{Stojkovi\ifmmode~\acute{c}\else \'{c}\fi{} and
  Valls}(1995)}]{Valls1}
\bibinfo{author}{\bibfnamefont{B.~P.}
  \bibnamefont{Stojkovi\ifmmode~\acute{c}\else \'{c}\fi{}}} \bibnamefont{and}
  \bibinfo{author}{\bibfnamefont{O.~T.} \bibnamefont{Valls}},
  \bibinfo{journal}{Phys. Rev. B} \textbf{\bibinfo{volume}{51}},
  \bibinfo{pages}{6049} (\bibinfo{year}{1995}).

\bibitem[{\citenamefont{\ifmmode \check{Z}\else
  \v{Z}\fi{}uti\ifmmode~\acute{c}\else \'{c}\fi{} and Valls}(1997)}]{Valls2}
\bibinfo{author}{\bibfnamefont{I.}~\bibnamefont{\ifmmode \check{Z}\else
  \v{Z}\fi{}uti\ifmmode~\acute{c}\else \'{c}\fi{}}} \bibnamefont{and}
  \bibinfo{author}{\bibfnamefont{O.~T.} \bibnamefont{Valls}},
  \bibinfo{journal}{Phys. Rev. B} \textbf{\bibinfo{volume}{56}},
  \bibinfo{pages}{11279} (\bibinfo{year}{1997}).

\bibitem[{\citenamefont{Halterman and Valls}(2000)}]{Valls3}
\bibinfo{author}{\bibfnamefont{K.}~\bibnamefont{Halterman}} \bibnamefont{and}
  \bibinfo{author}{\bibfnamefont{O.~T.} \bibnamefont{Valls}},
  \bibinfo{journal}{Phys. Rev. B} \textbf{\bibinfo{volume}{62}},
  \bibinfo{pages}{5904} (\bibinfo{year}{2000}).

\bibitem[{\citenamefont{Yuan et~al.}(2017)\citenamefont{Yuan, He, and
  Law}}]{Yuan}
\bibinfo{author}{\bibfnamefont{N.~F.~Q.} \bibnamefont{Yuan}},
  \bibinfo{author}{\bibfnamefont{W.-Y.} \bibnamefont{He}}, \bibnamefont{and}
  \bibinfo{author}{\bibfnamefont{K.~T.} \bibnamefont{Law}},
  \bibinfo{journal}{Phys. Rev. B} \textbf{\bibinfo{volume}{95}},
  \bibinfo{pages}{201109} (\bibinfo{year}{2017}).

\end{thebibliography}

\pagebreak

\onecolumngrid
\appendix

\section{Character on the Fermi surface: Dirac nodes, Weyl nodes, Majorana nodes}

The four different phases that appear in the phase diagram of the pseudoscalar and vector order parameters coupled to a magnetization 
order parameter have a peculiar character on the Fermi surface. By writing the gap matrix as $\Delta={\bf d}_{\bf k}\cdot{\bf s}$, the character 
on the Fermi surface can be addressed by studying the bulk spectrum
\begin{equation}
E_\pm({\bf k})=\sqrt{(\epsilon_{\bf k}-\mu)^2+|{\bf d}_{\bf k}|^2\pm|{\bf d}_{\bf k}\times{\bf d}^*_{\bf k}|},
\end{equation}
on the Fermi surface $\epsilon_{\bf k}=\mu$. With the vector ${\bf d}_{\bf k}=\chi {\bf k}+\boldsymbol{\psi}\times{\bf k}$ one can write the gap 
in terms of the condensate magnetization ${\bf \Sigma}_1$ and ${\bf \Sigma}_2$ as
\begin{equation}
\Delta_\pm({\bf k})=\sqrt{(|\chi|^2+|\boldsymbol{\psi}|^2)k^2-|\boldsymbol{\psi}\cdot{\bf k}|^2\pm k\sqrt{|{\bf \Sigma}_1|^2k^2-
|{\bf \Sigma}_1\cdot{\bf k}|^2+|{\bf \Sigma}_2\cdot{\bf k}|^2}},
\end{equation}
with ${\bf k}$ on the Fermi surface, $k=1$.  
For $\boldsymbol{\psi}=0$ the phase is fully gapped, with $\Delta_\pm=|\chi|$

\subsubsection{Nematic state}

In the nematic phase one has $\chi$ and $\boldsymbol{\psi}$ both real, so that ${\bf \Sigma}_1={\bf \Sigma}_2=0$, and the gap reads 
\begin{equation}
\Delta^{\rm nem}_\pm({\bf k})=\sqrt{|\chi|^2+|\boldsymbol{\psi}|^2-|\boldsymbol{\psi}\cdot\hat{\bf k}|^2},
\end{equation}
so that the phase is fully gapped as long as $\chi\neq 0$, whereas for $\chi=0$ it has a two double degenerate nodes for $\hat{\bf k}\parallel\boldsymbol{\psi}$. 
These nodes represent Dirac points and can be gapped by hexagonal warping \cite{FuPRBR2014}.

\subsubsection{Chiral state}

In the chiral phase one has $\chi$ real and $\boldsymbol{\psi}=\psi({\bf u}+i{\bf v})$, 
with ${\bf u}$ and ${\bf v}$ orthogonal unit vectors. Let us first consider the case $\chi=0$. The gap then reads
\begin{equation}
\Delta_\pm({\bf k})=\sqrt{|\boldsymbol{\psi}|^2-|\boldsymbol{\psi}\cdot{\bf k}|^2\pm |{\bf \Sigma}_2\cdot{\bf k}|},
\end{equation}
It is clear that only the gap $\Delta_-$ can be zero on a given 
point $(\theta,\phi)$ of the Fermi surface. 
Due to SO(3) symmetry we can choose for simplicity $\boldsymbol{\psi}=\psi(1,i,0)$, so that by writing $\hat{\bf k}=(\sin\theta\cos\phi,\sin\theta\sin\phi,\cos\theta)$
the gap reads
\begin{equation}
\Delta^{\rm chi}_\pm({\bf k})\propto\psi(1\pm|\cos(\theta)|).
\end{equation}
One has a node on the north pole $\theta=0$ and a node on the south pole $\theta=\pi$  on the Fermi sphere. Each node represents a Weyl point 
with topological charge $C=\pm2$, with $C=+2$ at the north pole and $C=-2$ at the south pole. These nodes cannot be gapped unless nodes with opposite 
topological charge are brought into contact. 

We can see this in more details by expanding the Hamiltonian in the reduced subspace of the conduction band for small momentum ${\bf q}$ around 
the the nodal points. For $\chi=0$ these are the north and south pole ${\bf k}^\pm_F=(0,0,\pm k_F)$, and the Hamiltonian reads
\begin{equation}
H^{\rm chi}_{\pm,{\bf q}}=\left[\begin{array}{cccc}
\pm v_F q_z & 0 & -i\psi q_+ & \pm 2i\psi k_F\\
0 & \pm v_F q_z & 0 & i\psi q_+\\
i\psi q_- & 0 & \mp v_F q_z & 0\\
\mp 2i\psi k_F & -i\psi q_- & 0 &  \mp v_F q_z 
\end{array}\right],
\end{equation}
where $q_\pm=q_x\pm i q_y$. We see that the Hamiltonian splits into two Weyl sub-blocks coupled by a mass term $m=2\psi k_F$ and the resulting 
eigenvalues give rise to two gapped bands at $\pm m$ and two gapless bands. Projecting onto the gapless states we find
\begin{equation}
h^{\rm chi}_{\pm,{\bf q}}=\pm\left[v_Fq_z\sigma_z-\frac{\psi^2}{m}(iq_+^2\sigma_+-iq_-^2\sigma_-)\right],
\end{equation} 
that is linearly dispersing along $q_z$ but quadratically dispersing along $q_x$ and $q_y$. One can show that the topological charge of these band 
crossing is $C=\pm 2$. 

When $\chi\neq 0$ the gap reads
\begin{equation}
\Delta^{\rm chi}_\pm({\bf k})\propto\psi\sqrt{(\chi/\psi)^2+1+\cos^2(\theta)\pm2\sqrt{\cos^2(\theta)((\chi/\psi)^2+1)+(\chi/\psi)^2\cos^2(\phi)\sin^2(\theta)}}.
\end{equation}
One can look for nodal solutions of $\Delta_-$, that reduces to solve $\cos^2\theta_W=1-(\chi/\psi)^2e^{\pm2i\phi}$, and find the there exist 
Weyl nodes with topological charge $|C|=1$ for $0<\chi<\psi$ only for $\phi=0,\pi$.  The Weyl node with topological charge 2 is separated into two 
Weyl nodes with topological charge $C=1$ at finite angles $\pm \theta_W$ in the $x,z$ plane ($\phi=0$), and analogously for the nodes at the 
south pole. The quadratic crossing splits into two linear crossing with $C=1$ in the north hemisphere and two linear crossing with $C=-1$ in the 
north hemisphere It follows that by increasing $\chi$ one moves the Weyl nodes toward the equator and for $\chi=\psi$ one has that Weyl nodes 
of opposite charge are brought into contact and split, so that for $\chi>\psi$ the system is fully gapped.

In the plane $\phi=0$ the nodes are located at $\sin\theta_W=\pm \chi/\psi$. We expand the Hamiltonian around the point 
$\hat{\bf k}_F=(\sin\theta_W,0,\cos\theta_W)$, and define radial and tangential momentum ${\bf q}=(q_{\parallel,x},q_{\parallel,y},q_\perp)$,

\subsubsection{TRSB 1 state}

In the TRSB phase 1 characterized by ${\bf \Sigma}_1\neq 0$ and ${\bf \Sigma}_2=0$ and one has $\chi$ real and $\boldsymbol{\psi}=i\psi{\bf n}$, 
with ${\bf n}$ a real unit vector. The gap reads
\begin{equation}
\Delta_\pm({\bf k})=\sqrt{(|\chi|^2+|\boldsymbol{\psi}|^2)-|\boldsymbol{\psi}\cdot\hat{\bf k}|^2\pm \sqrt{|{\bf \Sigma}_1|^2-
|{\bf \Sigma}_1\cdot\hat{\bf k}|^2}},
\end{equation}
Choosing ${\bf n}=(1,0,0)$ the gap then reads
\begin{equation}
\Delta_\pm^{\rm TRSB 1}\propto\left|\chi\pm \psi\sqrt{1-\sin^2(\theta)\cos^2(\phi)}\right|.
\end{equation}
For $\chi=0$ one obtain Dirac nodes at $\theta=\pi/2$, $\phi=0,\pi$ as for the nematic case. For $0<\chi<\psi$ one has nodal lines. These are best 
seen by choosing the coordinate in momentum space so to align the $z$ direction to the nematic director (that is by choosing ${\bf n}=(0,0,1)$) so that the gap reads
\begin{equation}
\Delta_\pm^{\rm TRSB 1}\propto\left|\chi\pm \psi\sin(\theta')\right|,
\end{equation}
with $\theta'$ the polar angle with respect to the $x$ axis. It is then clear that the Dirac point at $\chi=0$ evolves in a circle.

\subsubsection{TRSB 2 state}

Finally we now address the character on the Fermi surface of the gap in the TRSB 2 phase, where both ${\bf \Sigma}_1$ and ${\bf \Sigma}_2$ are non-zero 
but with ${\bf \Sigma}_2$ not maximal. In this case one can in general write $\boldsymbol{\psi}=(\cos(\alpha/2),i\sin(\alpha/2),0)$ and take $\chi$ real. The gap 
function in this case is not particularly enlightening. Nevertheless, one can show that for $0<\chi<\psi$ in general one has 2 Weyl points of topological charge 
$C=1$ in the north hemisphere and 2 Weyl points of topological charge $C=-1$ negative in the south hemisphere.  As in the purely chiral state 
the $\chi$ component moves the position of the Weyl points toward the equator and at $\chi=\psi$ they merge and split, so that for $\chi>\psi$ the state is gapped.

\section{Thermodynamic critical field in the TRSB phases}

As we pointed out in the main text, a crucial point for the existence of a TRSB phase with a non-zero condensate a dopants spin magnetization is that the total spin 
magnetization $M_{\rm s}$ be smaller than the thermodynamic critical field, $M_{\rm s}<H_{\rm cr}$. The latter can be calculated by the condensation energy, that 
is the free energy evaluated in the minimum at the mean-field value of the order parameters. The case of the condensation of the vector order parameter only is 
particularly simple and the value of the thermodynamic field has a simple form that allows us to study the condition for TRSB. We present here the derivation of the ratio 
$M_{\rm s}/H_{\rm cr}$ for this particular case and results may be extended straightforwardly for the other TRSB phases presented in the main text.

It is rather reasonable to assume the coupling $c_2<0$, according to which the dopants and condensate spin magnetization tend to align along a given direction. The GL free energy then reads
\begin{equation}
F=a_3M^2+a_2\psi^2+(b_2+b_2')\psi^4+c_2M\psi^2,
\end{equation}   
where $\psi>0$ is the absolute value of the condensate order parameter and $M$ the absolute value of the dopants magnetization. At the minimum one has $M_0=-\frac{c_2}{2a_3}\psi_0^2$, that is positive under the assumption of $a_3>0$ and $c_2<0$, and $\psi_0^2=-2a_2a_3/(4a_3(b_2+b_2')-c^2_2)$, that is positive under the assumption that $a_2<0$ and $4a_3(b_2+b_2')-c^2_2>0$. These two condition are essential for the stability of the superconducting phase described by a GL free energy up to forth order. The thermodynamic critical field is then given by
\begin{equation}
H_{\rm cr}=\sqrt{-8\pi F[M_0,\psi_0]}.
\end{equation}
Analogously, the value of the total spin magnetization is written as $M_{\rm s}=M_0+\mu \Sigma_2=(-\frac{c_2}{2a_3}+\mu)\psi_0^2$. The ratio between the the total spin magnetization and the critical field is then written as
\begin{equation}
\frac{M_{\rm s}}{H_{\rm cr}}=\sqrt{\frac{a_3}{2\pi}}\frac{\mu-c_2/(2a_3)}{\sqrt{4a_3(b_2+b_2')-c_2^2}}.
\end{equation}
For a paramagnetic system $a_3>0$ is temperature independent in the range of temperature of interest and we have that $M_{\rm s}/H_{\rm cr}$ 
is temperature independent. Furthermore, a stable superconducting phase is stabilized by a large $b_2$, so that the ratio $M_{\rm s}/H_{\rm cr}$ is smaller than one for sufficiently large $b_2$.

\section{Microscopic Theory of Superconductivity in Bi$_2$Se$_3$}

In the main text we studied superconductivity in the odd parity channel for a SO(3) Dirac Hamiltonian and we referred to Bi$_2$Se$_3$ 
as a possible material system. The Bi$_2$Se$_3$ family is well described by the 3D massive Dirac equation Eq.~(\ref{H0}) that, with the 
construction of the Dirac matrices in terms of spin ${\bf s}$ and $p_z$ orbital $\boldsymbol{\sigma}$ Pauli matrices given in Table \ref{tab}, 
can be casted in the form of a Dirac Hamiltonian. The actual point group of the material is $D_{3d}$ and we now specify to this case. 

We now consider the full interacting problem described by purely interlayer interaction, since it is assumed that they play a major role. We go a 
step beyond the purely local interaction discussed in  Ref.~\cite{FuBerg} and extend the attraction to nearest neighbors. 
In the Cooper channel the interaction reads
\begin{equation}
H_{\rm int}=-\frac{1}{2}\sum_{i\neq j}\sum_{{\bf k},{\bf k}';s,s'}V({\bf k}-{\bf k}')c^\dag_{{\bf k},i,s}c^\dag_{-{\bf k},j,s'}c_{-{\bf k}',j,s'}c_{{\bf k}',i,s},
\end{equation}
with $V({\bf q})$ the Fourier transform of the interaction potential. A detailed microscopic description of the nearest neighbor interaction in 
Bi$_2$Se$_3$ is beyond the scope of the present work and we simply assume that an expansion at lowest order in ${\bf q}={\bf k}-{\bf k}'$ can 
be done. We take into account the anisotropy along the $z$-direction typical of the material by splitting the momentum as ${\bf k}=({\bf k}_\parallel,k_z)$ 
and introducing effective length scales $a$ and $a_z$ on order of the lattice constants. Defining ${\bf k}=({\bf k}_\parallel,k_z)$
the interaction reads
\begin{equation}
V({\bf k},{\bf k}')=\frac{V}{2}\left(1+a^2{\bf k}_\parallel\cdot{\bf k}'_\parallel+a_z^2k_zk_z'\right).
\end{equation} 
These terms can involve only vectorial representations and tends to increase the 
strength of channel interaction. The next step consists in expanding the interaction in irreducible representations of the point group $D_{3d}$. 
When SO(3) is broken down to $D_{3d}$ the vector order parameter splits as $\boldsymbol{\psi}\to(\boldsymbol{\psi}_\parallel,\psi_z)$ and 
we can define the following basis functions
\begin{eqnarray}\label{basis-functions}
\Gamma_x^1({\bf k})&=&-i\gamma^5\gamma^2k_z,\qquad \Gamma_x^2({\bf k})=-i\gamma^5\gamma^3k_y,\\
\Gamma_y^1({\bf k})&=&-i\gamma^5\gamma^3k_x,\qquad \Gamma_y^2({\bf k})=-i\gamma^5\gamma^1k_z,\nonumber\\
\Gamma_z({\bf k})&=&-i\gamma^5(\gamma^1k_y-\gamma^2k_x),\nonumber
\end{eqnarray} 
where $\Gamma_x^{1,2}$ and $\Gamma_y^{1,2}$ belong to $E_u$ and $\Gamma_z$ belongs to $A_{2u}$. 
Following \cite{SigristUeda} and focusing on the odd-parity sector we write 
the gap matrix as
\begin{equation}
\hat{\Delta}=\chi \gamma^5+\boldsymbol{\psi}\cdot\boldsymbol{\gamma}+a\psi_zF_y+\psi_x(a_zF_x^1-aF_x^2)++\psi_y(aF_y^1-a_zF_y^2),
\end{equation}
with both the pseudo-scalar $\chi$ and the vector $\boldsymbol{\psi}$ order parameters. We see that the extra terms contains the contraction of 
the momentum with the pseudo-vector $\gamma^5\boldsymbol{\gamma}$, that is the possible odd-parity term involving the momentum only allowed 
by symmetry, as explained in the next section. Setting the chemical potential in the conduction band, $\mu>m$, upon projecting onto the conduction band, one obtains the gap matrix
\begin{equation}
\Delta_{\bf k}=\chi\tilde{\bf k}\cdot\tilde{\bf s}+\boldsymbol{\psi}\times\tilde{\bf k}\cdot\tilde{\bf s}(1+\mu a/v)
\end{equation}
for the isotropic case $a=a_z$. For the anisotropic case $a\neq a_z$, the projection of the basis functions Eq.~(\ref{basis-functions}) onto the 
conduction band produces the basis function introduced in Ref.~\cite{Venderbos2015}, and by introducing the parameters $\lambda=(1+\mu a/v)$ 
and $\lambda_z=(1+\mu a_z/v_z)$ the gap matrix reads
\begin{equation}
\hat{\Delta}=\chi\tilde{\bf k}\cdot\tilde{\bf s}+\lambda\psi_z(\tilde{s}_x\tilde{k}_y-\tilde{s}_y\tilde{k}_x)+
\psi_x(\lambda_z\tilde{s}_x\tilde{k}_z-\lambda\tilde{s}_z\tilde{k}_y)+\psi_y(\lambda\tilde{s}_z\tilde{k}_x-\lambda_z\tilde{s}_x\tilde{k}_z),
\end{equation}
where the momentum has been rescaled as $\tilde{\bf k}=(vk_x,vk_y,v_zk_z)/\mu$. We see that the nearest neighbor interaction 
rescales the momentum only of the vector channel.

We now consider the role of magnetic impurities. In the normal phase Nb-doped Bi$_2$Se$_3$ is found to be paramagnetic 
\cite{Qiu}, so that we do not consider direct ferromagnetic coupling between the magnetic dopants. Assuming that the dopants 
couple in the same way to the spin of the two orbitals, the Zeeman coupling reads
\begin{equation}
H_{\rm Z}=-\sum_{i}J_i\int d{\bf r}~s_i({\bf r})m_i({\bf r}),
\end{equation}
where ${\bf m}({\bf r})$ is the magnetic moment density of the dopants, ${\bf s}_{s,s'}({\bf r})=
\sum_ic^\dag_{i,s}({\bf r})c_{s',i}({\bf r}){\bf s}_{s,s'}$ is the 
electron spin operator, and $J_x=J_y=J\neq J_z$ are the anisotropic Zeeman coupling constants. The spin operator ${\bf s}=(s_x,s_y,s_z)$ does not 
transform as a pseudovector according to the transformation rules of SO(3) dictated by the representations of the $\gamma$-matrices in Tab.~\ref{tab}. 
Indeed, it is evident from Tab.~\ref{tab} that it is constructed with the components of ${\bf S}_\parallel\equiv\gamma^5\boldsymbol{\gamma}$ and 
${\bf S}_\perp\equiv\gamma^0\gamma^5\boldsymbol{\gamma}$, which represent generalized spin operator of the bonding and anti-bonding 
configurations of the two orbitals.  Considering only the ${\bf q}=0$ component of the magnetization we can then write the Zeeman coupling as
\begin{equation}
H_{\rm Z}=-J S^1_\parallel M^1-J S^2_\parallel M^2-J_z S^3_\perp M^3.
\end{equation}
This coupling breaks the SO(3) symmetry by mixing the two operators ${\bf S}_\parallel$ and ${\bf S}_\perp$. By projecting the Zeeman term onto 
the eigenstates of the conduction band at ${\bf k}=0$ one has $H_{\rm Z}=-\tilde{\bf s}\cdot \hat{J} \cdot{\bf M}$, with 
$\hat{J}={\rm diag}(J,J,J_z)$, where the projection of both ${\bf S}_i$ generalized spin operator gives the spin of 
the conduction band $\tilde{\bf s}$. For ${\bf k}$ on the Fermi surface one has the mapping
\begin{equation}
S_\parallel^i\to\tilde{s}_i-\frac{\mu}{m+\mu}\tilde{k}_i\tilde{\bf k}\cdot\tilde{\bf s},\qquad 
S_\perp^i\to\frac{m}{\mu}\tilde{s}_i+\frac{\mu}{m+\mu}\tilde{k}_i\tilde{\bf k}\cdot\tilde{\bf s}.
\end{equation}

\begin{center}
\begin{table}
\begin{tabular}{|c||c|c|c|c|c|c|c|}
\hline
 &$\gamma^0$ & $\gamma^5$ & $\gamma^0 \gamma^5$ & $\vec \gamma$ &  $\gamma^0 \vec \gamma$ & $\gamma^0 \gamma^5 \vec \gamma$ & $\gamma^5 \vec \gamma$ \\ \hline
Fu model & $\sigma_x$ & $\sigma_y s_z$ & $\sigma_z s_z$ & (-$\sigma_y s_y,\sigma_y s_x,\sigma_z)$ & $(\sigma_z s_y,-\sigma_z s_x,\sigma_y)$ & $(\sigma_x s_x,\sigma_x s_y,s_z)$ & $(s_x,s_y,\sigma_x s_z)$  \\ \hline
I &+&-&-&-&-&+&+  \\ \hline
T &+&+&-&+&-&-&-  \\ \hline
C &+&+&-&+&-&-&-  \\ \hline
$M_x$ &+&-&-&(-,+,+)&(-,+,+)&(+,-,-)&(+,-,-)  \\
\hline
\end{tabular}
\caption{Classification of Dirac algebra matrices, their realization in the Fu model and their symmetry properties. I stands for inversion symmetry, T for time-reversal symmetry, C for charge conjugation, and $M_x$ is the mirror about the $yz$ plane. From these, there are a pseudo-scalar $\gamma^0\gamma^5$ and a vector $\gamma^0\gamma^i$ that are odd under parity and, combined with the momentum, give rise to even parity pairing, thus only correcting the momentum-independent even-parity channels.  The remaining two pseudo-vectors $\gamma^0\gamma^i$ and $\gamma^0\gamma^5\gamma^i$ are even under parity and combined with the momentum thay can give odd parity pairing. 
}\label{tab}
\end{table}
\end{center}

\section{Derivation of the Ginzburg - Landau free energy}

We now derive the Ginzburg-Landau free energy starting from the microscopic model. For simplicity we refer to the isotropic case $a_z=a$, $v_z=v$, 
but keep the anisotropy in the Zeeman term. The inclusion of the Zeeman coupling  to the Bogolyubov-deGennes Hamiltonian in the Nambu basis 
$\Psi_{\bf k}=({\bf c}_{\bf k},{\cal T}{\bf c}_{\bf k})^T$ results in the addition of a term $H_Z$ with equal sign for electrons and holes. 
We can now integrate away the fermionic degrees of freedom and obtain a non-linear functional for the order parameters, 
\begin{equation}
{\cal S}=\int_0^\beta d\tau\frac{1}{V}{\rm Tr}\left[\hat\Delta^\dag\hat\Delta\right]-\frac{1}{\beta}{\rm Tr}\ln({\cal G}_0^{-1}-\Sigma),
\end{equation}
with $-{\cal G}_0^{-1}=\partial_\tau+(H_0-\mu)\tau_z$ and $\Sigma=\tau^+\hat{\Delta}+\tau_+H_{\rm Z}$, 
and the trace is over all the degrees of freedom, ${\rm Tr}\equiv T\sum_\omega\int d{\bf k}$. As usual, the microscopic GL theory is 
obtained by expanding the non-linear action in powers of the fields,
\begin{equation}\label{FunctionalDelta}
-\frac{1}{\beta}{\rm Tr}\ln(-{\cal G}_0^{-1}+\Sigma)=-\frac{1}{\beta}{\rm Tr}\ln(-{\cal G}_0^{-1})-
\frac{1}{\beta}\sum_{n=1}^\infty\frac{1}{n}{\rm Tr}({\cal G}_0\Sigma)^n.
\end{equation}
We first focus on the superconducting order parameter and set $J=0$. The second order terms are given by 
$\langle\Delta_{\bf k}\Delta_{\bf k}^\dag\rangle_{(2)}$ and the forth order coefficient are determined by the forth order averages 
$\langle\Delta_{\bf k}\Delta_{\bf k}^\dag\Delta_{\bf k}\Delta_{\bf k}^\dag\rangle_{(4)}$, 
where $\langle\ldots \rangle_{(2)}=T\sum_{\omega_n}\int \frac{d{\bf k}}{(2\pi)^3}G_+G_-{\rm Tr}[\ldots]$ and 
$\langle\ldots \rangle_{(4)}=\frac{T}{2}\sum_{\omega_n}\int \frac{d{\bf k}}{(2\pi)^3}G_+^2G_-^2{\rm Tr}[\ldots]$, 
with the unperturbed Green's function given by $G_\pm=(i\omega_n\mp\xi_{\bf k})^{-1}$, $\xi_{\bf k}=\epsilon_{\bf k}-\mu$ 
and $\epsilon_{\bf k}=\sqrt{v^2{\bf k}^2+m^2}$ is the dispersion of the conduction band.

The matrix which describes the gap function in spin space for the two component representation $\boldsymbol{\psi}_\parallel$ is 
\begin{equation}
\Delta_{\bf k}=\chi\tilde{\bf k}\cdot{\bf s}+\sum_{i=x,y}\psi_i{\bf d}_i\cdot{\bf s},
\end{equation}
with ${\bf d}_x=(0,-\tilde{k}_z,\tilde{k}_y)\lambda$, ${\bf d}_y=(\tilde{k}_z,0,-\tilde{k}_x)\lambda$. The coefficients of the GL free energy of the second order 
couplings are given by
\begin{eqnarray}
a_1&=&\frac{1}{V}-\chi_0(T)\langle \tilde{k}^2\rangle_{\rm FS},\\
a^{ij}_2&=&\frac{1}{V}-\chi_0(T)\langle {\bf d}_i\cdot{\bf d}_j\rangle_{\rm FS}(1+\mu a/v)^2
\end{eqnarray}
where $\chi_0(T)=N(\epsilon_F)\int d\epsilon \tanh(\epsilon/2T)/\epsilon$, $N(\epsilon_F)=\mu^2\sqrt{1-m^2/\mu^2}/(2\pi^2v^3)$ is the density of states at the Fermi level, $\langle\ldots\rangle_{\rm FS}=\int \frac{d{\bf k}}{(2\pi)^3}\delta(\epsilon_{\bf k}-\mu)\ldots$ stands for Fermi surface average, 
and the coupling of the components vector order parameter are diagonal, $a^{ij}_2=\delta_{ij}a_2$. The second order coefficients 
allows us to determine the critical temperature of the independent channels, and we find
\begin{eqnarray}
\frac{1}{V}&=&\chi_0(T_\chi)(1-m^2/\mu^2),\\
\frac{1}{V}&=&\frac{2}{3}(1+\mu a/v)^2\chi_0(T_\psi)(1-m^2/\mu^2).
\end{eqnarray}
It becomes clear that nearest neighbor interactions can increase the critical temperature of the vector order parameter, 
so that it is reasonable to consider both at the same time and study the coupled theory. 

The coefficient $a_3$ of the second order term in ${\bf M}$ contains two terms: i) the susceptibility of the free magnetic moment, and 
ii) the term coming from the second order expansion Eq.~(\ref{FunctionalDelta}), and it can be approximated to a positive constant. 

The higher order terms in the GL free energy are obtained by the higher order expansion of the functional Eq.~(\ref{FunctionalDelta}). 
The third order term gives the coupling between the magnetization and the pseudo-vector ${\bf \Sigma}_1$ and ${\bf \Sigma}_2$ introduced in the main text. 
The anisotropic Zeeman breaks the SO(3) symmetry down to the $D_{3d}$ point group. By writing the third order coupling as
\begin{equation} 
F^{(3)}=ic_1{\bf M}_\parallel\cdot(\chi\boldsymbol{\psi}^*_\parallel-\chi^*\boldsymbol{\psi}_\parallel)+ic^z_2M_z\boldsymbol{\psi}_\parallel\times\boldsymbol{\psi}^*_\parallel
\end{equation}
the values of the coupling for the in-plane $c_2$ and out-of-plane $c^z_1$ components of the magnetization reads
\begin{eqnarray}
c^z_2&=&\frac{2}{3}J_z(1-m^2/\mu^2)\mu \kappa,\\
c_1&=&\frac{4}{3}J(1-m^2/\mu^2)\mu \kappa.
\end{eqnarray}
The coefficient of the fourth order terms  for the isotropic case become
\begin{equation}
b_1=\kappa(1-m^2/\mu^2)^2,\qquad
b_2=\frac{8}{15}\lambda^4b_1,\qquad
b_2'=\frac{4}{15}\lambda^4b_1,\qquad
d_1=\frac{4}{3}\lambda^2b_1,\qquad
d_2=\frac{2}{3}\lambda^2b_1
\end{equation}
with $\kappa=N(\epsilon_F)7\zeta(3)/(8(\pi T_c)^2)$. It follows that the phase diagram for the anisotropic case governed by the couplings 
$J$ and $J_z$ is qualitatively similar to the one in the main text.

\section{Parametrization of the vector $\boldsymbol{\psi}$}
\label{parametrization}

We now present a parametrization of the vector order parameter that allows to simplify the analysis of the free energy of the coupled system. 
The order parameter $\boldsymbol{\psi}=(\psi_x,\psi_y,\psi_z)$ is described by three complex or six real degrees of freedom. If we write 
$\boldsymbol{\psi} = \psi_0 ({\bf u} + i {\bf v})$ with $u^2 + v^2 = 1$ with $u = |{\bf  u}|$ and $v = |{\bf v}|$, the different 
terms in the free energy take the form
\begin{align}
\boldsymbol{\psi} \cdot \boldsymbol{\psi}^*&= \psi_0^2,\\
\boldsymbol{\psi} \times \boldsymbol{\psi}^*&= -i\psi_0^2 2 {\bf u} \times {\bf v} \\
\boldsymbol{\psi} \cdot \boldsymbol{\psi}&= \psi_0^2(u^2-v^2 + 2i {\bf u} \cdot {\bf v}) = \psi_0^2 \sqrt{1-(2{\bf u} \times {\bf v})^2} e^{i \phi} \\
\phi &= \arctan \frac{2{\bf u} \cdot {\bf v}}{u^2-v^2}
\end{align}
This motivates the parametrization $u = \cos \alpha/2$, $v=\sin \alpha/2$, $0 \leq \alpha \leq \pi$ and 
\begin{align}
X &= 2 {\bf u} \cdot {\bf v} = 2uv \cos \theta = \sin \alpha \cos \theta \\
Y &= 2 |{\bf u} \times {\bf v}| = 2uv \sin \theta = \sin \alpha \sin \theta \\
Z &= u^2-v^2 = \cos \alpha
\end{align}
where the variables are so labeled due to the resemblance to spherical coordinates. $\theta$ is defined as the relative angle 
between ${\bf u}$ and ${\bf v}$. When we consider the coupling to the magnetization, the absolute directions of ${\bf u}$ and ${\bf v}$ need to be defined. The simplest way is to define $\phi'$ and $\theta '$ as the absolute angles in spherical coordinates of the unit vector ${\bf u}\times{\bf v}/uv$ and $\gamma'$ as the absolute azimuthal angle of ${\bf u}$ with respect to the axis ${\bf u}\times{\bf v}/uv$. The six real variables that parametrize $\boldsymbol{\psi}$ are therefore $\psi, \alpha, \theta, \phi', \theta', \gamma'$.

The two TRSB phases discussed in the text can be distinguished by the way rotation symmetry is broken in each of them. In the TRSB 1 phase, where only $\boldsymbol{\Sigma}_1$ is finite, the ground state remains invariant under SO(2) rotations around the $\boldsymbol{\Sigma}_1$ axis. In the phases where $\boldsymbol{\Sigma}_2$ is finite, assuming that $\boldsymbol{\Sigma}_2$ points in the z direction, the vector order parameter is given by  $\boldsymbol{\psi} = \psi_0 [u(\cos \gamma',\sin \gamma',0)+iv(\cos(\gamma'-\theta),\sin(\gamma'-\theta),0)]$. In the fully chiral phase where $u=v=1/\sqrt{2}$ and $\theta=\pi/2$, we have $\boldsymbol{\psi} = \psi_0 e^{i\gamma'}(1,i,0)/\sqrt{2}$, so that a rotation around the $\boldsymbol{\Sigma}_2$ axis corresponds to a shift in $\gamma'$, which becomes a phase shift of $\boldsymbol{\psi}$. This phase shift is not a pure gauge because of the presence of $\chi$, but if we shift the phase of $\chi$ by the same amount, this operation becomes a true symmetry of the fully chiral phase. Indeed, both $\boldsymbol{\Sigma}_1$ and $\boldsymbol{\Sigma}_2$ remain invariant under this mixed gauge-rotational symmetry. Finally, this symmetry is broken in the hybrid phase TRSB 2, where $u\neq v$, and no residual rotation symmetry remains. 

\subsection{Hybrid TRSB solution of $F_{\chi,\boldsymbol{\psi}}$ }

We now consider in detail the coupling between the scalar and the vector phase in the case the SO(3) is broken down to $D_{3d}$. 
The GL free energy is given by
\begin{equation}
F=F_\chi+F_\psi+F_{\chi,\psi}. 
\end{equation}
The phase digram as a function of temperature and interaction 
parameters $b_2'$, $d_1$, and $d_2$ that admits three possible phases: i) the $A_{1u}$ phase, where only the scalar $\chi$ 
condenses, $\chi=\chi_A$ and $\boldsymbol{\psi}=0$, ii) a nematic time reversal invariant phase with $\chi=\chi_{\rm N}$ and 
$\boldsymbol{\psi}=\boldsymbol{\psi}_{\rm N}$ real, and iii) a hybrid TRSB phase with $\chi=\chi_{\rm h}$ and 
$\boldsymbol{\psi}=\boldsymbol{\psi}_{\rm h}$ complex. We define $T_A$ the condensation temperature of the scalar $A_{1u}$ 
phase and $T_E$ the condensation temperature of the two-component $E_{u}$ and assume the system to be at $T<T_E$.

Employing the parametrization introduced in the previous section for the vector $\boldsymbol{\psi}$ and setting $\chi = \chi_0 e^{i\gamma}$ 
the free energy is written as 
\begin{equation}\label{FcpFull}
F = a_1 \chi_0^2 + b_1 \chi_0^4 + a_2 \psi_0^2 +(b_2 +b_2'Y^2)\psi_0^4  + [d_1+2d_2-2d_2 \sqrt{1-Y^2} \cos (\phi-2\gamma)]\chi_0^2 \psi_0 ^2
\end{equation}
with $\phi = \arctan X/Z$. Note that $\phi$ and $Y$ can be taken as independent variables since they also parametrize the full sphere, so 
that we can minimize independently for $Y$ and $\phi$ without a constraint. Since the parametrization of $\boldsymbol{\psi}$ contains 4 real 
parameters, it does contain arbitrary changes of the overall phase (i.e. gauge transformations), so that in principle we can assume $\chi$ to 
be real and $\gamma=0,\pi$. However, when studying vortices or configurations where the phase changes in real space we need to keep 
$\gamma$.

The usefulness of this parametrization when $\gamma=0$ is that $\phi$ can always be minimized independently, since it is clear that regardless 
of the rest of the parameters one obtains a lower energy by setting $\phi = 0,\pi$ for positive or negative $d_2$. This corresponds to having 
$X/Z = \tan \alpha \cos \theta = 0$, which gives two options. First, if $\alpha = 0,\pi$, then either $u$ or $v$ is zero, which is a nematic solution 
with the same phase as $\chi$, hence no TRSB phase. Second, if $\theta=\pm \pi/2$ then ${\bf u}$ and ${\bf v}$ are orthogonal and this is a TRSB  
phase, where the relative weight of $u$ and $v$ is obtained from minimizing with respect to $Y = \sin \alpha$ (for finite alpha since otherwise we 
are in the previous solution). 

The minimization with respect to $Y$ now has the following options. If $b_2'>0$ and $d_2>0$, then we always get $Y=0$ and a nematic phase, since both terms 
that contain $Y$ want it to be as small as possible. If $b_2'<0$ and $d_2>0$ then there is a competition between $b_2'$ which favors the chiral solution 
and $d_2$ which favors the nematic solution. The value of $Y$ is obtained from 
\begin{equation}
2b_2'\psi_{\rm h}^4 Y + 2d_2 Y\psi_{\rm h}^2\chi_{\rm h}^2/\sqrt{1-Y^2} = 0
\end{equation}
which gives the solution 
\begin{equation}
Y = \sqrt{1-\frac{d_2^2}{(b_2')^2}\frac{\chi_{\rm h}^4}{\psi_{\rm h}^4}}, 
\end{equation}
which interpolates between nematic and the standard chiral as $Y$ goes from 0 to 1. Finally, if $b_2'<0$ and $d_2<0$ one has $Y=0$ and 
$\phi-2\gamma=\pi$, that corresponds to a solution in which $\boldsymbol{\psi}$ is real and $\chi=-i\chi_0$.

\end{document}